\documentclass[showpacs,preprintnumbers,amsmath,amssymb,letterpaper,twocolumn]{revtex4}
\usepackage{graphicx}
\usepackage{bm}
\usepackage{amssymb, latexsym, amsmath, textcomp}

\newcommand{\ra}{\rangle}
\newcommand{\la}{\langle}
\renewcommand{\v}[1]{\boldsymbol{#1}}
\newcommand{\prt}{\partial}
\newcommand{\ie}{{\it ie~}}

\begin{document}

\title{
Dichotomy in underdoped high $T_c$ superconductors and\\ 
spinon-dopon approach to $t$-$t'$-$t''$-$J$ model} 
\author{Ying Ran}
\author{Xiao-Gang Wen}
\homepage{http://dao.mit.edu/~wen}
\affiliation{Department of Physics, Massachusetts Institute of Technology,
Cambridge, Massachusetts 02139}
\date{November 1, 2006}

\begin{abstract} 
We studied underdoped high $T_c$ superconductors using a spinon-dopon approach
(or doped-carrier approach) to $t$-$t'$-$t''$-$J$ model, where spinon carries
spin and dopon carries both spin and charge. In this approach, the mixing of
spinon and dopon describes superconductivity. We found that a nonuniform mixing
in $k$-space is most effective in lowering the $t'$ and $t''$ hopping energy.
We showed that at mean-field level, the mixing is proportional to quasiparticle
spectral weight $Z_-$.  We also found a simple monte-carlo algorithm to
calculate $Z_{-}$ from the projected spinon-dopon wavefunction, which confirms
the mean-field result.  Thus the non-uniform mixing caused by $t'$ and $t"$
explains the different electron spectral weights near the nodal and anti-nodal
points ({\it i.e.} the dichotomy) observed in underdoped high $T_c$
superconductors.  For hole-doped sample, we found that $Z$ is enhanced in the
nodal region and suppressed in the anti-nodal region. 
For electron doped sample, the same approach leads to a suppressed $Z$ in the
nodal region and enhanced in the anti-nodal region, in agreement with
experimental observations.
\end{abstract} 
\pacs{71.10.-w, 74.72.-h, 74.25.Jb}
\maketitle

\section{Introduction}

One powerful experimental technique to study high-$T_c$ material is the Angular
Resolved Photoemission Spectroscopy (ARPES)\cite{RevModPhys.75.473}. ARPES
study for the pseudogap region showed a strong anisotropy of the electron
spectral function in momentum space\cite{97043617370,04278242766}. Basically it
was found that in the nodal direction, excitations are more quasi-particle
like; while in the anti-nodal direction, excitations have no quasi-particle
peak.  This is the so-called dichotomy. If one lowers the temperature to let
the material to go into superconducting phase, it was found that anti-nodal
direction also has a small quasi-particle peak. Tunneling experiments show that
the underdoped samples are
inhomogeneous\cite{01456717214,PhysRevB.64.100504,mcelroy:197005}. Due to this
inhomogeneity, it is possible that the underdoped sample can be separated into
optimal doped regions and underdoped regions, and the quasi-particle peak only
comes from the optimal doped region. With such a point of view, it is possible
that even the superconducting phase can have a very anisotropic electron
spectral function in momentum space. 

Exact diagonalization on $t$-$t'$-$t''$-$J$ model ($t'$ and $t''$ stand for
next nearest neighbor and next next nearest neighbor, respectively) with 32
sites has been done\cite{PhysRevB.52.R15711,PhysRevB.56.6320} for hole-doped
case (one hole doped). It was found that if $t'=t''=0,J=0.3t$, then the
quasiparticle weight $Z_-$ is almost a constant along the direction
$(\pi,0)$-$(0,\pi)$: $Z_-=0.311$ at $(\pi/2,\pi/2)$, and $Z_-=0.342$ at
$(\pi,0)$. However if one put in $t'=-0.3t,t''=0.2t,J=0.3t$, which is an
optimal parameter fitting for $\mbox{Sr}_2\mbox{CuO}_2\mbox{Cl}_2$, then there
is a strong dichotomy feature: $Z_-=0.353$ at $(\pi/2,\pi/2)$, and $Z_-=0.029$
at $(\pi,0)$. This suggests that the dichotomy can be a result of $t'$ and
$t''$ hopping. 

Exact diagonalization was also done for the electron doped case (a few electrons
doped on 32 sites)\cite{PhysRevB.56.6320}, where $Z_+$ was measured. Due
to the particle-hole symmetry at half-filling, we know that if $t'=t''=0$,
$Z_+$ and $Z_-$ are equal up to a momentum shift of $(\pi,\pi)$. Therefore
$Z_+$ are also flat along the direction $(\pi,0)$-$(0,\pi)$ in pure $t$-$J$ model.
But when we put in $t'=-0.3t,t''=0.2t,J=0.3t$, the particle-hole symmetry was
broken. $Z_+$ was found to develop a strong anisotropy along
$(\pi,0)$-$(0,\pi)$: $Z_+=0.005$ at $(\pi/2,\pi/2)$, and $Z_+=0.636$ at
$(\pi,0)$.

What mechanism can destroy the quasi-particle coherence in the anti-nodal
region? The simplest thing comes into one's mind is that we need some other
things to destroy it. For example, neutron scattering experiments indicate that
there are some low energy magnetic
fluctuations\cite{99084762357,95072773972,98014012369,PhysRevB.57.6165}, and it
was proposed\cite{97043617370,04278242766} that magnon scattering process
can destroy the quasi-particle coherence in anti-nodal region. 
In this paper, however, we propose a physically different scenario: The
dichotomy is due to the $t'$ and $t''$ hopping terms.  The quasiparticle
spectral weight $Z_k$ is naturally suppressed in some region in $k$-space to
lower the $t'$ and $t"$ hopping energy.  This contradicts a naive thinking that
hopping always enhance $Z_k$.  Using the $t$-$t'$-$t''$-$J$ model, we will show
that the new scenario can explain the distribution of $Z_k$ for both hole-doped
and electron-doped samples in a unified way.

If we believe that the dichotomy is driven by the $t'$ and $t''$ hopping terms,
then there is an important issue:  Is there a mean-field theory and the
corresponding trial wavefunction that captures this mechanism?  

One way to understand high-$T_c$ superconductors is to view them as doped
Mott-insulators. Under Zhang-Rice singlet mapping\cite{3155883}, the minimal
model which includes the essential Mott physics is $t$-$J$ model on square
lattice. 
On the analytical side, a powerful mean-field theory for $t$-$J$ model, the
slave-boson approach, was developed\cite{2986272,3246071,9015906}. This
approach emphasizes the fractionalization picture of the doped Mott insulator:
electron is splited into a spinon (a fermion with spin and no charge) and a
holon (a boson with charge and no spin), which characterize the low energy
excitations of the doped Mott insulator. This mean-field approach also
successfully predicted the pseudogap metal for underdoped samples. On the
numerical side, the same physics picture gives rise to the projected BCS
wavefunction\cite{3206105}(pBCSwf), which turns out to be a very good trial
wavefunction for $t$-$J$ model.  However, more detailed studies of
pBCSwf\cite{8880404,bieri-2006} indicate that the slave-boson approach fail to
explain the dichotomy. So a momentum dependent quasiparticle weight
$Z_{k}$ remains to be a big challenge for slave-boson theory.


In this paper, we will use a new spinon-dopon approach\cite{05449451727} and
the corresponding trial wavefunction to study the underdoped samples.  Instead
of using spinons and holons, in the new approach, we use the spinons
and the bond states of spinons and holons to describe the low energy
excitations. The  bond states of spinons and holons  are called dopons which
are charge-$e$ spin-$\frac12$ fermions.  The spinon-dopon approach leads to a
new trial wavefunction, the projected spinon-dopon wave function (pSDwf).  The
new trial wave function turns out to be an improvement over the old projected
BCS wavefunction (pBCSwf). 


The holon condensation in slave-boson approach correspond to spinon-dopon
mixing. However, in the spinon-dopon approach, the mixing can have a momentum
dependence, which is beyond the mean-field slave-boson approach. 
If we set the mixing to have no momentum dependence,
then the pSDwf turns out to be identical to the old pBCSwf. So the pSDwf is a
generalization of the pBCSwf.

Now the question is, why the mixing wants to have strong momentum dependence?
The answer is that the wavefunction with momentum dependent mixing can make the
hopping more coherent, and therefore gain hopping energy. Roughly speaking, the
pSDwf with momentum dependent mixing is the summation of the old projected BCS
wavefunction together with hopping terms $c_i^{\dagger}c_j$ acting on it. Here
one should notice that the old pBCSwf, with uniform mixing, already has a
pretty good $t$ hopping energy. But to have a good $t'$ and $t''$ hopping
energy, the mixing needs to have a momentum dependence, along the direction
from $(\pi,0)$ to $(0,\pi)$. Our Monte Carlo calculation shows that pSDwf with
momentum dependent mixing is indeed a better trial wavefunction in energetic
sense.  To get a quantitative sense how big is the improvement, we find that
the energy of a doped hole in pSDwf is about $0.4t$ lower than that of  a doped
hole in pBCSwf.  This is a very big improvement, indicating that the
spin-charge correlation (or more precisely, the spin configuration near a doped
hole) is much better described by pSDwf than pBCSwf.

Can one measure the momentum dependence of mixing? The answer is yes. In this
article we will show that the mixing is directly related to $Z_-$, the
quasi-particle weight, which is measurable in ARPES. Roughly speaking, mixing
is proportional to $Z_-$. We have also developed a Monte Carlo technique to
calculate $Z_-$. The calculation shows that momentum dependent mixing pSDwf
indeed has strong anisotropy in momentum space, and consistent with the
observed dichotomy.

Comparing pBCSwf and pSDwf, we like to point out two wave functions have
similar background spin-spin correlation and similar spin energy.  However,
that pBCSwf does not capture the detailed charge dynamics.  The new trial
wavefunction, pSDwf, contains more correct spin-charge correlation. As a
result,  the energy of doped holes/electrons is much lower in the pSDwf.  The
holes/electrons in the pSDwf reproduce the correct momentum dependence of
quasi-particle spectral weight.  We also expect our pSDwf to have a strong
momentum dependence in quasi-particle current, which may explain the
temperature dependence of superfluid density of High $T_c$
superconductors\cite{liang:117001}.

\section{Spinon-dopon Approach and pSDwf}

\subsection{Slave-boson Approach and Projected BCS Wavefunction --
Why the approach fails to capture $k$-dependent features?}\label{SBA} 

Why would we want to introduce the spinon-dopon
approach to $t$-$J$ model? Let us firstly look into the previous mean
field approach, more specifically, slave-boson approach. The general $t$-$J$
model can be written in terms of electron operator: 
\begin{align}
H_{tJ}=&J\sum_{\la ij\ra\in
NN}\left(\mathbf{S}_i\cdot\mathbf{S}_j-\frac{1}{4}n_i n_j\right)\notag\\
&-\sum_{ij}t_{ij}\mathbf{P}\left(c_i^{\dagger}c_j+c_j^{\dagger}c_i\right)\mathbf{P}.
\end{align} 
Here the projection operator $\mathbf{P}$ is to ensure the
Hamiltonian is acting within the physical Hilbert space: one each site, the
physical states are $\vert\uparrow\rangle$, $\vert\downarrow\rangle$ or $\vert
0\rangle$, i.e., no double occupancy.

Slave-boson approach\cite{2986272,3246071} emphasizes the spin-charge
separation picture. In that approach, one splits
electron operator into spinon and holon operators: 
\begin{align} 
c_{i\sigma}=f_{i\sigma}b_i^{\dagger},
\end{align} 
where $f$ is spinon, carrying spin $1/2$ and charge $0$, $i$ labels site and
$\sigma$ labels spin; $b$ is holon, carrying spin $0$ and charge $1$.  This
splitting enlarges the Hilbert space. To go back to physical Hilbert space, a
local constraint is needed: 
\begin{align}
f_{i\uparrow}^{\dagger}f_{i\uparrow}+f_{i\downarrow}^{\dagger}f_{i\downarrow}+b_i^{\dagger}b_i=1
\label{sbconst} .
\end{align} 

Due to spin interaction, spinons form a $d$-wave paired state. The
superconducting phase is realized through an additional holon condensation at
momentum $k=0$.  Within such a construction, the quasi-particle weight $Z$ is
proportional to doping $x$ everywhere in k-space, in both nodal and anti-nodal
region. To see this, one can simply look at the mean-field Green function of
electron: 
\begin{align}
\label{sbz}
\langle c_k^{}c^{\dagger}_k\rangle=\langle b^{\dagger}_{k=0}b_{k=0}^{} f_k^{}
f^{\dagger}_k\rangle=x\langle f_k^{}f^{\dagger}_k\rangle .
\end{align} 
Therefore $x$ is the residue of quasi-particle pole and $Z=x$ is
independent of $k$.

Slave-boson approach is supposed to capture the physics of spin-charge
separation. It has successfully generated the phase diagram of High-$T_c$
superconductor. But this approach, at least at mean-field level, could not
capture some more detailed features, such as momentum dependence of
quasi-particle weight or the quasi-particle current. One can argue that
including gauge fluctuation, those detailed features may be reproduced, but
here we will try to develop another approach which can capture these features
at mean-field level.

Before we go into the new spinon-dopon approach, let us see how far one can go
using slave-boson approach. One can actually try to build a trial wavefunction
based on slave-boson mean-field approach. We know that the mean-field approach
enlarged the Hilbert space, and the resulting wavefunction lies outside the
physical Hilbert space. Only when one includes the full gauge fluctuations can
one go back to the physical Hilbert space. 

So one way to include  the full gauge fluctuations, is to build the mean-field
ground state first, and then do a projection from the enlarged Hilbert space to
the physical Hilbert space. The wavefunction after projection would serve as a
trial wavefunction for the physical Hamiltonian. This projected wavefunction is
supposed to incorporate the effect of gauge fluctuation of the slave-boson
approach, and may answer the question that, after including gauge fluctuation,
whether slave-boson approach can capture the detailed features like dichotomy. 

The mean-field ground state for underdoped case can be constructed as follows.
Let $N_h$ be the number of holes, $N_f$ is the number of spinons, and
$N=N_h+N_f$ is the total number of sites. The slave-boson
mean-field ground state is then given by
\begin{align}
\vert\Phi_{SB,mean}\rangle=(b_{k=0}^{\dagger})^{N_h}\prod_{k}(u_k+v_k
f_{k\uparrow}^{\dagger}f_{-k\downarrow}^{\dagger})\vert0\rangle ,
\end{align}
where the spinon part of the wavefunction is a standard $d$-wave pairing state:
\begin{align}
\frac{v_k}{u_k}=\frac{\Delta(k)}{\xi_k+\sqrt{\xi_k^2+\Delta(k)^2}}\label{d-wave},
\end{align} 
where 
\begin{align} 
\xi_k&=-2\chi(\cos k_x +\cos k_y )-\mu   \nonumber \\
\Delta(k)&=\Delta(\cos k_x -\cos k_y )&&\mbox{(d-wave)}.
\label{dwave} 
\end{align}
Here $\mu$ is the chemical potential to give the correct average number of
spinon $\langle\sum_{i}f_{i\sigma}^{\dagger}f_{i\sigma}\rangle=N_f$; $\chi$ and
$\Delta$ are mean-field parameters which have been found to be
$\frac{\Delta}{\chi}=2$ \cite{4279595} at half-filling, and
$\frac{\Delta}{\chi}$ decreases to zero at doping around $J/t$.

Now one can do a projection to go back to the physical Hilbert space. The
constraint for physical Hilbert space is Eq.(\ref{sbconst}). This constraint
ensures that the total number of spinon must be $N_f$ and there is no double
occupancy of spinon: spinon number $n_{f,i}$ at site $i$ has to be either 0 or
1. One can easily see that the resulting wavefunction is the usual Projected
$d$-wave BCS Wavefunction (pBCSwf): 
\begin{align} 
\vert\Phi_{PBCS}\rangle&=
P_D^{SB} P_N^{SB}\vert\Phi_{SB,mean}\rangle\\ &=P_D P_N \prod_{k}(u_k+v_k
c_{k\uparrow}^{\dagger}c_{-k\downarrow}^{\dagger})\vert0\rangle\\ &\propto
P_D\left( \sum_k a(k)
c_{k\uparrow}^{\dagger}c_{-k\downarrow}^{\dagger}\right)^{N_f/2}\vert0\rangle,
\end{align} 
where in the first line, $P_N^{SB}$ is the projection into fixed
total number of particles, i.e., $N_h$ holons and $N_f$ spinons; while
$P_D^{SB}$ is the projection into physical Hilbert space, i.e., removing all
states not satisfying constraint Eq.(\ref{sbconst}). In the second line, $P_N$
is the projection into fixed total number of electrons, which has to be $N_f$,
$P_D$ is the projection which removes all double occupancies. $a(k)$ is defined
as $a(k)=\frac{v_k}{u_k}$. 

Projected BSC wavefunction turned out to be a surprisingly good trial
wavefunction for $t$-$J$ model\cite{3206105}. However numerical
studies\cite{8880404,bieri-2006} showed that the quasi-particle weight is
almost a constant along the direction from $(\pi,0)$ to $(0,\pi)$, i.e., it fails
to reproduce the dichotomy. The quasi-particle current of pBCSwf is also pretty
smooth in the $k$ space\cite{8880404}.  It is because pBCSwf is unable to
capture the momentum dependence properties that we need a new approach to
underdoped high $T_c$ superconductors.

\subsection{How to capture $k$-dependence features?\\
 -- Spinon-dopon approach and projected spinon-dopon wavefunction} 

Rebeiro and Wen\cite{05449451727} developed this new mean-field approach
trying to capture the spinon-holon recombination physics. In the following we
briefly review their work. We know that at low temperature, spinon and holon
recombine pretty strongly to give electron-like quasi-particle. So it is
natural to introduce dopon operator -- a bound state between a spinon and a
holon -- to describe low energy excitations.  Note that a dopon has the same
quantum number as an electron and describes a doped electron (or hole).  But
the Mott and spin-liquid physics at half filling should also be addressed. So
one should also keep the spinon operator. As a result, two types of fermions
are introduced here: spinon $f$ and dopon $d$. Spinon carries spin $1/2$ and no
charge, and dopon carries spin $1/2$ and charge $1$. By introducing these two
types of fermions one enlarges the Hilbert space: now there are 16 states per
site, among them only three are physical. The three physical states on site $i$
can be represented in terms of spinon and dopon fermions as: 
\begin{align}
\vert\uparrow\rangle=\vert\uparrow_f\rangle, 
&&\vert\downarrow\rangle=\vert\downarrow_f\rangle,
&&\vert0\rangle=\frac{1}{\sqrt{2}}
\vert\uparrow_f\downarrow_d-\downarrow_f\uparrow_d\rangle  .
\label{SDHS}
\end{align} 
Here please notice that the constraints are two-fold: firstly there
must be one $f$ spinon per site, secondly $d$ dopon has to form a local singlet
with the spinon.

One can do a self-consistent mean-field study. The mean-field Hamiltonian takes
the form: 
\begin{align} 
H_{mean}&=(-2\chi(\cos k_x+\cos
k_y)-\mu)f_{k\alpha}^{\dagger}f_{k\alpha}^{}\notag\\ &+\Delta(\cos k_x-\cos
k_y)f_{k\uparrow}^{\dagger}f_{-k\downarrow}^{\dagger}+\epsilon_k
d_{k\alpha}^{\dagger}d_{k\alpha}^{}\notag\\ &+\beta_k
f_{k\alpha}^{\dagger}d_{k\alpha}^{}+h.c. .
\label{SDMH} 
\end{align} 
Here $H_{mean}$ can be divided into three parts: spinon part, dopon part and
spinon-dopon interaction. The spinon part describes the usual $d$-wave paired
ansatz: $\chi=J\langle f_{k\alpha}^{\dagger}f_{k\alpha}^{}\rangle$,
$\Delta=J\langle f_{k\uparrow}^{}f_{-k\downarrow}^{}\rangle$. The dopon part is
simply a free dopon band, with $\epsilon_k$ determined by high energy ARPES
measurement. Note that $\epsilon_k$ is not taken as tunable mean-field
parameter. Finally the spinon-dopon interaction is described by a $k$-dependent
hybridization, roughly speaking $\beta_k=\epsilon_k \la
d_i^{\dagger}f_i^{}\ra$. One can see that $d_i^{\dagger}f_i^{}$ is a bosonic
field carrying charge $1$ and spin $0$. Its non-zero average value corresponds
to holon condensation in slave-boson approach, which leads to
superconductivity. $\mu$ is the chemical potential required to tune the doping.

Along this line Rebeiro and Wen did a mean-field phase diagram, and
successfully fit to ARPES data and tunneling
data\cite{PhysRevB.74.155113,PhysRevLett.97.057003}. Here we try to emphasize
that the main lesson we learned from this new mean-field approach is that one
can have a $k$-dependent hybridization at mean-field level (in Eq.(\ref{SDMH})
this hybridization is controlled by $\beta_k$ and energy spectrum of spinon
band and dopon band.), which is roughly the counterpart of holon condensation
in slave-boson approach. This is why one can study detailed features like
dichotomy in this new approach.

Several open questions naturally arise in this new approach. It seems there are
two types of excitations, spinon and dopon, what do they look like? We also
know that mean-field approach is not very reliable, so it would be nice to
understand the physical trial wavefunction corresponding to the new mean-field
approach, from where we would know exactly what we are doing. In the following
we try to answer these questions. 

Let us construct trial wavefunctions based on this spinon-dopon mean-field
approach. One can simply take a mean-field ground state wavefunction, then do a
projection back into the physical Hilbert space, just like the way we did in
the slave boson case: 
\begin{align}
\vert\Phi_{PSD}\rangle=P_{SD}P_N\vert\Phi_{SD,mean}\rangle  .
\end{align} 
Here $P_N$ is the projection into fixed number of spinon and dopon, which gives
the correct doping; and $P_{SD}$ is the projection into to physical Hilbert
space Eq.(\ref{SDHS}). $\vert\Phi_{SD,mean}\rangle$ is the ground state
wavefunction of some mean-field Hamiltonian in the form of Eq.(\ref{SDMH}).
Suppose we know how to do this projection numerically, one can do a variational
study of the these Projected Spinon-Dopon Wavefunctions (pSDwf), to see what is
the lowest-energy ansatz. In general, however, the full projection is not doable, so we
develop a simple numerical technique to do a local projection to have some
rough idea about what kind of wavefunction is energetically favorable (See Appendix
\ref{LP}). What we found is that the best trial wavefunction for underdoped
case has the following form: 
\begin{align}
\vert\Phi_{PSD}^{SC}\rangle&=P_{SD}P_{N}\vert\Phi_{SD,mean}^{SC}\rangle\\
&=P_{SD}P_{N}\exp\left(\sum_k
b(k)\tilde{f}_{k\uparrow}^{\dagger}\tilde{f}_{-k\downarrow}^{\dagger}\right)\vert0\rangle\\
&\propto P_{SD}\left(\sum_k
b(k)\tilde{f}_{k\uparrow}^{\dagger}\tilde{f}_{-k\downarrow}^{\dagger}\right)^{\frac{N+N_h}{2}}\vert0\rangle,
\end{align} 
where 
\begin{align}
\tilde{f}^{\dagger}_{k\alpha}=\sqrt{1-\beta_k^2}f_{k\alpha}^{\dagger}+\beta_k
d_{k\alpha}^{\dagger} .
\end{align} 
Here $\tilde{f}$ form a $d$-wave paired state and
the superscript $SC$ means this wavefunction is superconducting. $b(k)$
and $\beta_k$ are some real functions and we assume $\beta_k=\beta_{-k}$ to
respect time reversal symmetry. For this particular ansatz, full projection is
doable in low doping limit. In section \ref{NM} we develop the numerical method
to do the full projection and we will see that this wavefunction is a even
better trial wavefunction than pBCSwf.

Note that the total number of $f$ and $d$ fermions is
$N+N_h$. Also $P_{SD}$ requires one $f$-fermion per site, so totally $N$
$f$-fermions. Therefore we must have $N_h$ $d$-fermions, which gives the
correct doping.

\subsection{How does pSDwf capture the $k$-dependent features? -- properties
of wavefunction before projection $\vert\Phi_{SD,mean}^{SC}\rangle$: $Z_k$ at
mean-field level.} 

The form of $\vert\Phi_{PSD}^{SC}\rangle$ looks very similar to pBCSwf,
basically we are constructing a pairing wavefunction based on hybridized
fermion $\tilde{f}_k$. In the next section we will see that
$\vert\Phi_{PSD}^{SC}\rangle$ and pBCSwf are indeed closely related. For the
moment let us have a closer look at the wavefunction
$\vert\Phi_{SD,mean}^{SC}\rangle$ before projection. The idea is that physical
properties may not change drastically after the projection. In this case the
mean-field level understanding will give us insight of the wavefunction after
the projection. 

First of all it is obvious that this wavefunction is superconducting. That is
because the nonzero $\beta_k$ signals the mixing between spinon and dopon $\la
f_{k\alpha}^{\dagger}d_{k\alpha}\ra\neq0$, and thus signals breaking of charge
conservation. It is natural to believe the superconductivity survives after
projection.

Let us introduce the other combination of $f$ and $d$ fermions: 
\begin{align}
\tilde{d}^{\dagger}_{k\alpha}=-\beta_k f_{k\alpha}^{\dagger}+
\sqrt{1-\beta_k^2} d_{k\alpha}^{\dagger} ,
\end{align} 
and the quasi-particle operators: 
\begin{align}
\gamma_{k\uparrow}^{\dagger}&=\tilde{u}_k\tilde{f}_{k\uparrow}^{\dagger}-\tilde{v}_k\tilde{f}_{-k\downarrow}\\
\gamma_{-k\downarrow}&=\tilde{u}_k\tilde{f}_{-k\downarrow}+\tilde{v}_k\tilde{f}_{k\uparrow}^{\dagger},
\end{align} 
where 
\begin{align}
\tilde{u}_k=\frac{1}{\sqrt{1+b(k)^2}}&&\tilde{v}_k=\frac{b(k)}{\sqrt{1+b(k)^2}},
\end{align} 
are the coherent factors for a $d$-wave paired state.
We can show that
\begin{equation*}
\vert\Phi_{SD,mean}^{SC}\rangle=\exp\left(\sum_k
b(k)\tilde{f}_{k\uparrow}^{\dagger}\tilde{f}_{-k\downarrow}^{\dagger}\right)\vert0\rangle,
\end{equation*}
satisfies: 
\begin{align} 
\tilde{d}_{k\alpha}\vert\Phi_{SD,mean}^{SC}\rangle=0,
\end{align} 
and
\begin{align}
\gamma_{k\alpha}\vert\Phi_{SD,mean}^{SC}\rangle=0 .
\end{align} 
The mean-field
Hamiltonian which can generate $\vert\Phi_{SD,mean}^{SC}\rangle$ as ground
state is simply: 
\begin{align}
H_{mean}=\sum_k\left(\epsilon_{\tilde{f}}(k)\gamma_{k\alpha}^{\dagger}
\gamma_{-k\alpha}^{}+\epsilon_{\tilde{d}}(k)
\tilde{d}_{k\alpha}^{\dagger}\tilde{d}_{k\alpha}^{}\right),
\end{align} 
with $\epsilon_{\tilde{f}}(k),\epsilon_{\tilde{d}}(k)\geqslant0$.
Later we will see that there are physical reasons that
$\epsilon_{\tilde{d}}(k)>\epsilon_{\tilde{f}}(k)$, meaning $\tilde{f}$ band is
lowest energy excitation, and $\tilde{d}$ band is fully gapped,
$\epsilon_{\tilde{d}}(k)>0$ for any $k$.

We can express $f$ and $d$ fermions in terms of $\gamma$ and $\tilde{d}$ fermions:
\begin{align}
f_{k\alpha}&=\sqrt{1-\beta_k^2}(\tilde{u}_{k}\gamma_{k\alpha}+\tilde{v}_k
\epsilon_{\alpha\delta}\gamma_{-k\delta}^{\dagger})-\beta_k
\tilde{d}_{k\alpha}\label{MFf}\\
d_{k\alpha}&=\beta_k(\tilde{u}_{k}\gamma_{k\alpha}+\tilde{v}_k
\epsilon_{\alpha\delta}\gamma_{-k\delta}^{\dagger})
+\sqrt{1-\beta_k^2}\tilde{d}_{k\alpha}.
\label{MFd}
\end{align} 
Based on Eq.(\ref{MFf}) and (\ref{MFd}), it is easy to obtain:
\begin{align} 
\frac{\langle\Phi_{SD,mean}^{SC}\vert
f_{k\alpha}^{\dagger}f_{k\alpha}^{}\vert\Phi_{SD,mean}^{SC}\rangle}
{\langle\Phi_{SD,mean}^{SC}\vert\vert\Phi_{SD,mean}^{SC}\rangle}
&=(1-\beta_k^2)\tilde{v}_k^2\\
\frac{\langle\Phi_{SD,mean}^{SC}\vert
d_{k\alpha}^{\dagger}d_{k\alpha}^{}
\vert\Phi_{SD,mean}^{SC}\rangle}{\langle\Phi_{SD,mean}^{SC}
\vert\vert\Phi_{SD,mean}^{SC}\rangle}&=\beta_k^2\tilde{v}_k^2.
\end{align} 
We know that the mean-field wavefunction should give one
$f$-fermion and $x=\frac{N_h}{N}$ $d$-fermion per site on average:
\begin{align} \sum_{k}(1-\beta_k^2)\tilde{v}_k^2&=N\\
\sum_{k}\beta_k^2\tilde{v}_k^2&=N_h 
\end{align} 
In the low doping limit
$x\rightarrow0$, it is clear from the above relations that $\beta_k^2\propto
x$. 

Now let us understand how to calculate $Z_-$ and $Z_+$ on this mean-field
wavefunction. $Z_+$ and $Z_-$ are defined to be: 
\begin{align}
Z_{-,k}&=\frac{\vert\langle N-1,k \vert
c_k\vert\Phi_{GS}^N\rangle\vert^2}{\langle N-1,k\vert
N-1,k\rangle\langle\Phi_{GS}^N\vert\Phi_{GS}^N\rangle}\label{zdf-},\\
Z_{+,k}&=\frac{\vert\langle N+1,k \vert
c_k^{\dagger}\vert\Phi_{GS}^N\rangle\vert^2}{\langle N+1,k\vert
N+1,k\rangle\langle\Phi_{GS}^N\vert\Phi_{GS}^N\rangle}\label{zdf+} ,
\end{align}
where $\vert N-1,k\rangle$ ($\vert N+1,k\rangle$) are the lowest-energy $N-1$
($N+1$) electron states which have nonzero overlap with
$c_k\vert\Phi_{GS}^N\rangle$ ($c_k^{\dagger}\vert\Phi_{GS}^N\rangle$).

In our mean-field wavefunction, the lowerest energy excited states are given by
creating $\gamma_{k}$-quasi-particle. Note that now $d_{k}^{\dagger}$ is the
hole creation operator, so at mean-field level the $Z_k$ for spinon-dopon
wavefunction are: 
\begin{align}
Z_{-,k\uparrow}^{SD}&={\textstyle\frac{\vert\langle\Phi_{SD,mean}^{SC}\vert
\gamma_{k\uparrow}^{}d_{k\uparrow}^{\dagger}\vert\Phi_{SD,mean}^{SC}
\rangle\vert^2}{\langle\Phi_{SD,mean}^{SC}\vert\gamma_{k\uparrow}^{}
\gamma_{k\uparrow}^{\dagger}\vert\Phi_{SD,mean}^{SC}
\rangle\langle\Phi_{SD,mean}^{SC}\vert\Phi_{SD,mean}^{SC}\rangle}},
\notag\\
&=\beta_k^2\tilde{u}_k^{2}\label{zff-}\\
Z_{+,k\uparrow}^{SD}&={\textstyle\frac{\vert\langle\Phi_{SD,mean}^{SC}
\vert \gamma_{-k\downarrow}^{}d_{k\uparrow}\vert\Phi_{SD,mean}^{SC}\rangle
\vert^2}{\langle\Phi_{SD,mean}^{SC}\vert\gamma_{-k\downarrow}^{}
\gamma_{-k\downarrow}^{\dagger}\vert\Phi_{SD,mean}^{SC}\rangle\langle
\Phi_{SD,mean}^{SC}\vert\Phi_{SD,mean}^{SC}\rangle}}\notag\\
&=\beta_k^2\tilde{v}_k^2\label{zff+} .
\end{align}

At this moment, let us compare spinon-dopon wavefunction (SDwf) with BCS
wavefunction (BCSwf), both before projection (In Section \ref{RBPP} we will
compare them after projection). 

\begin{figure} 
\centerline{
\includegraphics[width=0.2\textwidth]{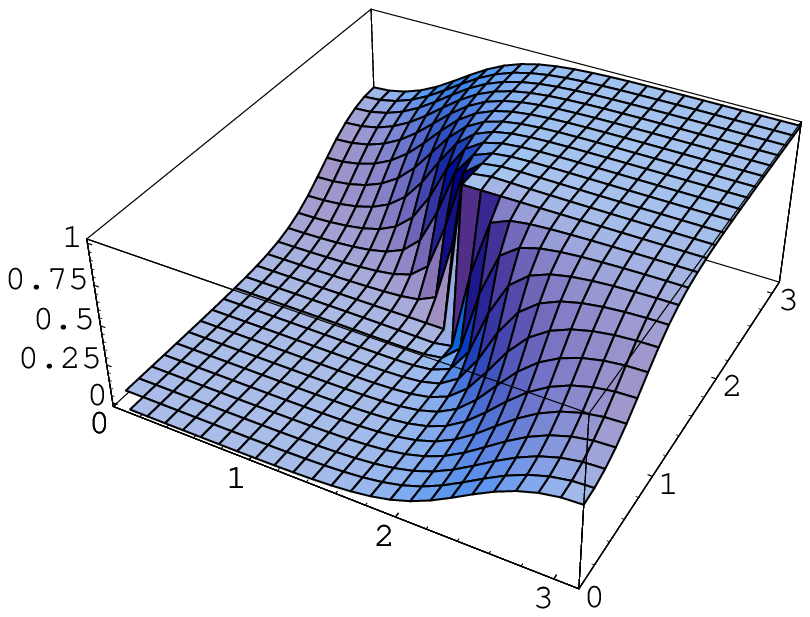}\;\;\;\;\includegraphics[width=0.2\textwidth]{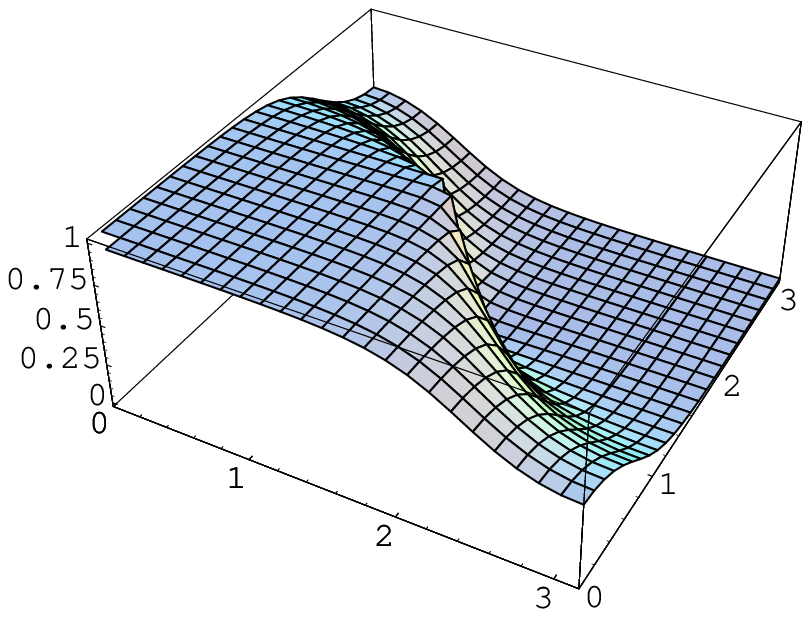}
} 
\caption{Plot of $u_k^2$ (left) and $v_k^2$ (right) within one quarter
Brillouin Zone, $k_x$ and $k_y$ range from $0$ to $\pi$.} \label{ukvk}
\end{figure}

In Section \ref{SBA} we view pBCSwf as the
projected slave-boson mean-field state into physical Hilbert space. 
We may also view pBCSwf as projected
BCSwf with all double occupancies removed:
\begin{align} 
\vert\Phi_{PBCS}\rangle&= P_D P_N\vert\Phi_{BCS}\rangle\\ &=P_D
P_N
\exp(\sum_{k}a(k)c_{k\uparrow}^{\dagger}c_{-k\downarrow}^{\dagger})\vert0\rangle\\
&\propto P_D\left( \sum_k a(k)
c_{k\uparrow}^{\dagger}c_{-k\downarrow}^{\dagger}\right)^{N_f/2}\vert0\rangle,
\end{align} 
where 
\begin{align}
\vert\Phi_{BCS}\rangle&=\exp(\sum_{k}a(k)c_{k\uparrow}^{\dagger}
c_{-k\downarrow}^{\dagger})\vert0\rangle\\
&\propto\prod_{k}(u_k+v_k
c_{k\uparrow}^{\dagger}c_{-k\downarrow}^{\dagger})\vert0\rangle .
\end{align} 
Before the projection, the spectral weight of the electron operator
$c_k$ can be calculated easily:
\begin{align}
Z_{-,k}^{BCS}=v_k^2=n_k&&Z_{+,k}^{BCS}=u_k^2=1-n_k\label{zbcs} ,
\end{align}
where $n_k=\la c_k^{\dagger}c_k^{}\ra_{BCS}$. For a $d$-wave BCSwf
Eq.(\ref{d-wave}), 
we can plot the $Z_k$ in Fig.\ref{ukvk}. In low doping limit, parameters are
taken as $\mu=0$, $\chi=1$, $\Delta=0.55$.  Such choice of parameters leads to
a pBCSwf with lowest average energy at half filling.

The $Z_{+,k}$ and $Z_{-,k}$ for the pBCSwf after the projection were also
calculated\cite{8880404}. Roughly speaking what was found is that the $Z_k$
profile after projection is similar to that before the projection.  There is a
quasi fermi surface, which is roughly along the diagonal direction, $Z_{-,k}$
is large inside the fermi surface and decreases very fast when you go outside
fermi surface; while $Z_{+,k}$ is large outside fermi surface, and decreases
fast when you go into fermi surface. But there is one big difference, which is
a reduction factor. For the $Z_{+,k}$, this reduction factor was found to be
proportional to $x$. But for $Z_{-,k}$, this reduction factor depends on $k$
and is finite (around $0.2$) for $k=(0,0)$ even at low-doping limit. From
slave-boson approach Eq.(\ref{sbz}), we already see that $Z\propto x$ at mean
field level. Basically at half filling, $Z=0$ and we have a Mott
insulator instead of a band insulator.

Notice that along diagonal direction $(\pi,0)-(0,\pi)$, the $Z_{k}^{BCS}$ is
dispersionless: $Z_{k}^{BCS}=0.5$, which does not have dichotomy feature; After
projection, there is a factor $x$ reduction, but $Z_{k}$ is still almost a
constant along the diagonal direction\cite{bieri-2006,8880404}.


To compare the calculated $Z$ from the BSCwf and SDwf, we note that 
the projected wave functions, pBCSwf and pSDwf, are closely related 
(see section \ref{RBPP}). More precisely: 
\begin{align}
&\vert\Phi_{PSD}^{SC}\rangle=\vert\Phi_{PBCS}\rangle\mbox{   if:}\notag\\
\tilde{u}_k=v_k,\;&\tilde{v}_k=u_k,\mbox{ and }\beta_k=\beta_0\mbox{
(constant)}\label{identify} 
\end{align} 
It is easy to understand this
identification at half-filling, since both wavefunctions simply give the same
spin-liquid (usually referred to as staggered flux spin liquid in literatures),
characterized by $\chi$ and $\Delta$. Now in the pSDwf $\tilde{f}=f$, with no
mixing with $d$-fermion. It is simply a particle-hole transformed pBCSwf, by
which $u_k$ transformed into $\tilde{v}_k$ and vice versa. The important
message is that  $u_k,v_k$ and $\tilde{u}_k,\tilde{v}_k$ characterize the spin
dynamics, but $\beta_k$ characterizes the charge dynamics. 

With this identification in mind, from Eq.(\ref{zff-},\ref{zff+}) and
(\ref{zbcs}) we immediately know that when $\beta_k=\beta_0$ these two wavefunctions give the same
mean-field $Z_k$ profile except that SDwf has an extra $x$ factor, because
$\beta_k^2=\beta_0^2\propto x$ in low doping limit.  However, when $\beta_k$
has a strong $k$-dependence, $Z_k$ from the two approaches can be very
different.

Let us think about whether or not these wavefunctions can capture dichotomy in
low doping limit. What did we learn from these mean-field result? We learned
that it is impossible to capture dichotomy by BCSwf, because in order to
capture the $k$-dependence along $(0,\pi)$--$(\pi,0)$, one has to tune $u_k,\
v_k$.  Because $d$-wave $u_k,v_k$ are constant along $(0,\pi)$--$(\pi,0)$, one
has to destroy the $d$-wave ansatz to have a $k$-dependent $u_k,\ v_k$ along
$(0,\pi)$--$(\pi,0)$.  This leads to a higher $J$ energy. On the other hand, it
is possible to capture dichotomy by SDwf, because one can tune $\beta_k$ to
have a strong $k$-dependence while keeping $\tilde{u}_k,\tilde{v}_k$ to
be $d$-wave ansatz. This will not destroy the spin background.  Based on our
experience of projection, we expect that even after projection, the above
statement is qualitatively true.

\subsection{Why mixing $\beta_k$ has a strong $k$-dependence? --Relation
between pBCSwf and pSDwf}\label{RBPP} 

In the last section we see that pSDwf can potentially capture the dichotomy
through a $k$ dependent $\beta_k$.
Now the issue is, why does the $\beta_k$ want to have a $k$ dependence that can
explain the dichotomy in $Z_k$?  Why does such $k$-dependent $\beta_k$ lead to a
pSDwf which is energetically more favorable? To understand this, we need to
know what a pSDwf looks like in real space.

The discussion below for identifying the relation between pSDwf and pBCSwf is
rather long. The result, however, is simple. Let us present the result here
first. We introduce $\tilde{\beta}_k=\beta_k/\sqrt{1-\beta_k^2}$. In low
doping limit, $\tilde{\beta}_k\approx \beta_k$. For the simplest one-hole case,
if $\tilde{\beta}_k$ has the simplest modulation in $k$-space
$\tilde{\beta}_k=\tilde{\beta}_0+2\tilde{\beta}_1(\cos k_x+\cos k_y)$, 
then the pSDwf can be viewed as pBCSwf mixed with the wavefunction generated by
the nearest neighbor hopping operators (see
Eq.(\ref{pfb1overlap1})). 
For more complicated
$\tilde{\beta}_k=\tilde{\beta}_0+2\tilde{\beta}_1(\cos k_x+\cos
k_y)+4\tilde{\beta}_2 \cos k_x\cos k_y+2\tilde{\beta}_3(\cos 2k_x +\cos 2k_y)$,
the pSDwf can be viewed as pBCSwf mixed with the wavefunction generated by
the nearest neighbor, next nearest neighbor and third
nearest neighbor hopping operators (see
Eq.(\ref{pfmh})).
Therefore to lower the hopping energy, finite $\beta_i$'s are naturally
developed. This is why $\beta_k$ with a proper $k$ dependence is more
energetically favorable.

Before we look into pSDwf, let us review what a pBCSwf looks like in real
space. One can do a Fourier transformation: 
\begin{widetext}
\begin{align}
\vert\Phi_{PBCS}\rangle=P_D\left( \sum_k a(k)
c_{k\uparrow}^{\dagger}c_{-k\downarrow}^{\dagger}\right)^{N_f/2}\vert0\rangle
=P_D\left( \sum_{\mathbf{R}_{i\uparrow},\mathbf{R}_{j\downarrow}}
a(\mathbf{R}_{j\downarrow}-\mathbf{R}_{i\uparrow})
c_{\mathbf{R}_{i\uparrow},\uparrow}^{\dagger}c_{\mathbf{R}_{j\downarrow},\downarrow}^{\dagger}\right)^{N_f/2}\vert0\rangle,
\end{align} 
where $a(\mathbf{r})=\sum_k a(k)\cos(k\cdot\mathbf{r})$. If we have
a spin basis $\{\mathbf{R}_{i\uparrow},\mathbf{R}_{j\downarrow}\}$, where
$\mathbf{R}_{i\uparrow}$ labels the positions of spin up electrons and
$\mathbf{R}_{j\downarrow}$ labels the positions of spin down electrons:
\begin{align}
\langle\{\mathbf{R}_{i\uparrow},\mathbf{R}_{j\downarrow}\}\vert\Phi_{PBCS}\rangle=
\left| 
\begin{array}{cccc}
a(\mathbf{R}_{1\downarrow}-\mathbf{R}_{1\uparrow})&a(\mathbf{R}_{1\downarrow}-\mathbf{R}_{2\uparrow})&\cdots&a(\mathbf{R}_{1\downarrow}-\mathbf{R}_{\frac{N_f}{2}\uparrow})\\
a(\mathbf{R}_{2\downarrow}-\mathbf{R}_{1\uparrow})&a(\mathbf{R}_{2\downarrow}-\mathbf{R}_{2\uparrow})&\cdots&a(\mathbf{R}_{2\downarrow}-\mathbf{R}_{\frac{N_f}{2}\uparrow})\\
\vdots&\vdots&\ddots&\vdots\\
a(\mathbf{R}_{\frac{N_f}{2}\downarrow}-\mathbf{R}_{1\uparrow})&a(\mathbf{R}_{\frac{N_f}{2}\downarrow}-\mathbf{R}_{2\uparrow})&\cdots&a(\mathbf{R}_{\frac{N_f}{2}\downarrow}-\mathbf{R}_{\frac{N_f}{2}\uparrow})
\end{array} 
\right| .
\end{align} 
We see that the overlap between a spin basis
and pBCSwf is simply a single slater determinant of a two-particle
wavefunction. This is why pBCSwf can be numerically simulated on a fairly large
lattice.

Now we go back to pSDwf. Up to a normalization constant, one can express pSDwf
as: 
\begin{align} 
\vert\Phi_{PSD}^{SC}\rangle =P_{SD}\left(\sum_k
b(k)(f_{k\uparrow}^{\dagger} +\tilde{\beta}_k
d_{k\uparrow}^{\dagger})(f_{-k\downarrow}^{\dagger}+\tilde{\beta}_k
d_{-k\downarrow}^{\dagger})\right)^{\frac{N+N_h}{2}}\vert0\rangle 
\end{align}
where $\tilde{\beta}_k=\beta_k/\sqrt{1-\beta_k^2}$. Since $\beta_k\propto
\sqrt{x}$, in the low doping limit, $\tilde{\beta}_k=\beta_k$.

One can also do a Fourier transformation into the real space: 
\begin{align} 
\vert\Phi_{PSD}^{SC}\rangle
=P_{SD}\left(\sum_{\mathbf{R}_{i\uparrow},\mathbf{R}_{j\downarrow}}
b(\mathbf{R}_{j\downarrow}-\mathbf{R}_{i\uparrow})(f_{\mathbf{R}_{i\uparrow}\uparrow}^{\dagger}
+\tilde{\beta}_0
d_{\mathbf{R}_{i\uparrow}\uparrow}^{\dagger}+\sum_{\mathbf{\delta}
}\tilde{\beta}_{\delta}
d_{\mathbf{R}_{i\uparrow}+\mathbf{\delta},\uparrow}^{\dagger})(f_{\mathbf{R}_{j\downarrow}\downarrow}^{\dagger}+\tilde{\beta}_0
d_{\mathbf{R}_{j\downarrow}\downarrow}^{\dagger}+\sum_{\mathbf{\delta}
}\tilde{\beta}_{\delta}
d_{\mathbf{R}_{j\downarrow}+\mathbf{\delta},\downarrow}^{\dagger}
)\right)^{\frac{N+N_h}{2}}\vert0\rangle \label{ffrs} ,
\end{align} 
where $\tilde{\beta}_{\mathbf{\delta}}$'s are the Fourier components of
$\tilde{\beta}_k$: 
\begin{align}
\tilde{\beta}_k=\tilde{\beta}_0+\tilde{\beta}_{\mathbf{x}}e^{ik_x}+\tilde{\beta}_{-\mathbf{x}}e^{-ik_x}+\tilde{\beta}_{\mathbf{y}}e^{ik_y}+\tilde{\beta}_{-\mathbf{y}}e^{-ik_y}+\cdots.
\end{align} 
We should only consider the rotation invariant $\tilde{\beta}_k$,
and let us only keep the first three Fourier components: 
\begin{align}
\tilde{\beta}_k=&\tilde{\beta}_0+2\tilde{\beta}_1(\cos k_x+\cos
k_y)+4\tilde{\beta}_2 \cos k_x\cos k_y\notag\\ &+2\tilde{\beta}_3(\cos 2k_x
+\cos 2k_y ) .
\end{align}

We claimed that if $\tilde{\beta}_k=\tilde{\beta}_0$, then pSDwf is identical
to pBCSwf if $b(k)=\frac{1}{a(k)}$. Let us see how that is true. Without
$\beta_{1,2,3}$, Eq.(\ref{ffrs}) is: 
\begin{align}
\vert\Phi_{PSD}^{SC}(\tilde{\beta}_0)\rangle=P_{SD}\left(\sum_{\mathbf{R}_{i\uparrow},\mathbf{R}_{j\downarrow}}
b(\mathbf{R}_{j\downarrow}-\mathbf{R}_{i\uparrow})(f_{\mathbf{R}_{i\uparrow}\uparrow}^{\dagger}
+\tilde{\beta}_0
d_{\mathbf{R}_{i\uparrow}\uparrow}^{\dagger})(f_{\mathbf{R}_{j\downarrow}\downarrow}^{\dagger}+\tilde{\beta}_0
d_{\mathbf{R}_{j\downarrow}\downarrow}^{\dagger})\right)^{\frac{N+N_h}{2}}\vert0\rangle\label{pfu}.
\end{align} 
\end{widetext}

What does a pBCSwf look like? If one does a particle-hole transformation
$c_{i\uparrow}^{\dagger}\rightarrow h_{i\downarrow}$, pBCSwf is: 
\begin{align}
&\vert\Phi_{PBCS}\rangle=P_D\left( \sum_k \frac{1}{a(k)}
h_{k\uparrow}^{\dagger}h_{-k\downarrow}^{\dagger}\right)^{\frac{N+N_h}{2}}\vert0\rangle\\
&=P_D\left( \sum_{\mathbf{R}_{i\uparrow},\mathbf{R}_{j\downarrow}}
b(\mathbf{R}_{j\downarrow}-\mathbf{R}_{i\uparrow})
h_{\mathbf{R}_{i\uparrow},\uparrow}^{\dagger}h_{\mathbf{R}_{j\downarrow},\downarrow}^{\dagger}\right)^{\frac{N+N_h}{2}}\vert0\rangle\label{pbh},
\end{align} 
where $P_D$ is the projection forbidding any empty site.

If we consider a spin basis
$\{\mathbf{R}_{i\uparrow},\mathbf{R}_{j\downarrow}\}$, with the empty sites
$\{\mathbf{R}_{k,0}\}$, then after particle-hole transformation, we have single
occupied sites $\{\mathbf{R}_{i\uparrow},\mathbf{R}_{j\downarrow}\}$, and
double occupied sites $\{\mathbf{R}_{k,0}\}$. So the position of spin up and
down sites in the hole representation are
$\{\mathbf{\tilde{R}}_{i\uparrow},\mathbf{\tilde{R}}_{j\downarrow}\}_h$, where
$\{\mathbf{\tilde{R}}_{i\uparrow}\}_h=\{\mathbf{R}_{i\uparrow}\}\cup\{\mathbf{R}_{k,0}\}$
and
$\{\mathbf{\tilde{R}}_{j\downarrow}\}_h=\{\mathbf{R}_{j\downarrow}\}\cup\{\mathbf{R}_{k,0}\}$.
The overlap of pBCSwf and the spin basis in hole representation is:
\begin{widetext} 
\begin{align}
&\langle\{\mathbf{R}_{i\uparrow},\mathbf{R}_{j\downarrow}\}\vert\Phi_{PBCS}\rangle=\langle\{\mathbf{\tilde{R}}_{i\uparrow},\mathbf{\tilde{R}}_{j\downarrow}\}_h\vert\Phi_{PBCS}\rangle \nonumber \\
=&\left| 
\begin{array}{cccc}
b(\mathbf{\tilde{R}}_{1\downarrow}-\mathbf{\tilde{R}}_{1\uparrow})&b(\mathbf{\tilde{R}}_{1\downarrow}-\mathbf{\tilde{R}}_{2\uparrow})&\cdots&b(\mathbf{\tilde{R}}_{1\downarrow}-\mathbf{\tilde{R}}_{\frac{N+N_h}{2}\uparrow})\\
b(\mathbf{\tilde{R}}_{2\downarrow}-\mathbf{\tilde{R}}_{1\uparrow})&b(\mathbf{\tilde{R}}_{2\downarrow}-\mathbf{\tilde{R}}_{2\uparrow})&\cdots&b(\mathbf{\tilde{R}}_{2\downarrow}-\mathbf{\tilde{R}}_{\frac{N+N_h}{2}\uparrow})\\
\vdots&\vdots&\ddots&\vdots\\
b(\mathbf{\tilde{R}}_{\frac{N+N_h}{2}\downarrow}-\mathbf{\tilde{R}}_{1\uparrow})&b(\mathbf{\tilde{R}}_{\frac{N+N_h}{2}\downarrow}-\mathbf{\tilde{R}}_{2\uparrow})&\cdots&b(\mathbf{\tilde{R}}_{\frac{N+N_h}{2}\downarrow}-\mathbf{\tilde{R}}_{\frac{N+N_h}{2}\uparrow})
\end{array} \right|.\label{pbd} 
\end{align} 
\end{widetext} 
The equation works
this way because if one simply expands the polynomial in Eq.(\ref{pbh}), each
sum will give you one term in the expansion of the determinant in
Eq.(\ref{pbd}), and Pauli statistics is accounted by the sign in determinant
expansion.

\begin{figure} \includegraphics[width=0.20\textwidth]{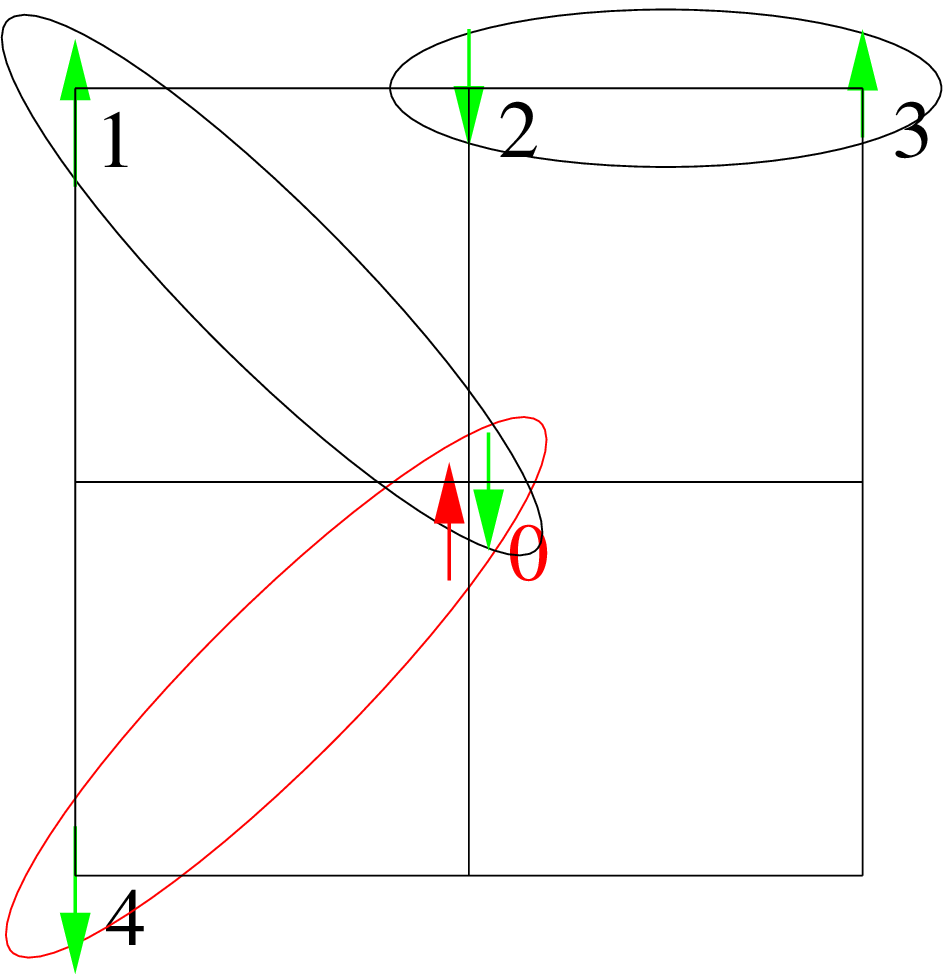}
\includegraphics[width=0.20\textwidth]{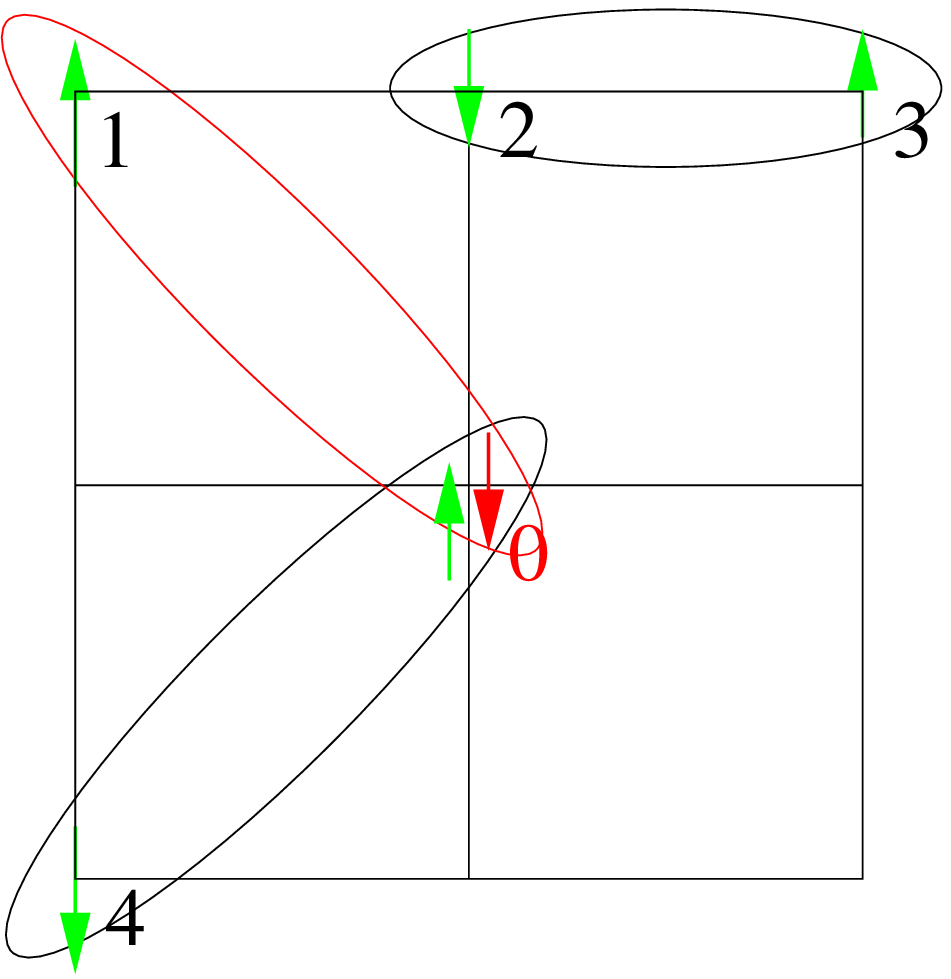} \caption{pSDwf with
only $\tilde{\beta}_0$. The site $0$ is empty. $f$-fermions are represented by
green spin, and $d$-fermion is represented by red spin. Black valence bonds are
bonds with of $f$-fermions, while red valence bond has a $d$-fermion. The two
figures are two contributions of the overlap between pSDwf  and a spin basis
$\vert1_{\uparrow}3_{\uparrow}2_{\downarrow}4_{\downarrow}0_{emp}\rangle$, and
they correspond to the same term in determinant Eq.(\ref{pfd}). The two figures
give rise to states: $ \mbox{left:
}\vert\uparrow_{1f}\downarrow_{0f}\uparrow_{3f}\downarrow_{2f}\uparrow_{0d}\downarrow_{4f}\rangle=\vert\uparrow_{1f}\uparrow_{3f}\downarrow_{2f}\downarrow_{4f}(\downarrow_{0f}\uparrow_{0d})\rangle$
and $ \mbox{right:
}\vert\uparrow_{1f}\downarrow_{0d}\uparrow_{3f}\downarrow_{2f}\uparrow_{0f}\downarrow_{4f}\rangle=-\vert\uparrow_{1f}\uparrow_{3f}\downarrow_{2f}\downarrow_{4f}(\uparrow_{0f}\downarrow_{0d})\rangle
$ The minus sign means that the two figures contribute additively.  }
\label{localmixing} 
\end{figure}

Now we can do the same analysis on pSDwf Eq.(\ref{pfu}). First of all
$\tilde{\beta}_0$ is not relevant in the wavefunction, since we are projecting
into a state with fixed number of $d$-fermion, which means that all
$\tilde{\beta}_0$ does is to give an overall factor $\tilde{\beta}_0^{N_h}$ in
front of the wavefunction. To have an overlap with spin basis
$\{\mathbf{R}_{i\uparrow},\mathbf{R}_{j\downarrow},\mathbf{R}_{k,0}\}$, we know
that on empty site $\mathbf{R}_{k,0}$ the expansion of polynomial
Eq.(\ref{pfu}) should give either $\vert\uparrow_f\downarrow_d\rangle$ or
$\vert\uparrow_d\downarrow_f\rangle$. After projection each case would
contribute to $\frac{1}{\sqrt{2}}\vert0\rangle$ where
$\vert0\rangle=\frac{1}{\sqrt{2}}(\vert\uparrow_f\downarrow_d\rangle+\vert\uparrow_d\downarrow_f\rangle)$

\begin{figure}
\includegraphics[width=0.20\textwidth]{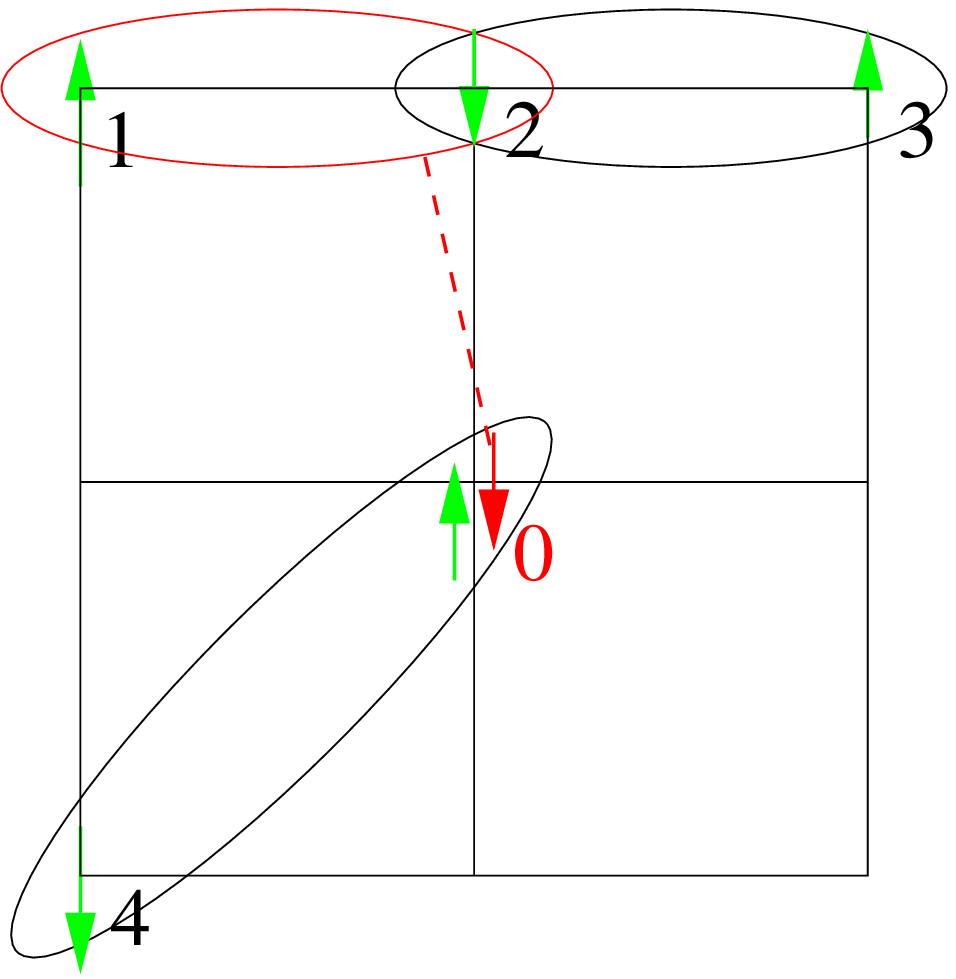}
\includegraphics[width=0.20\textwidth]{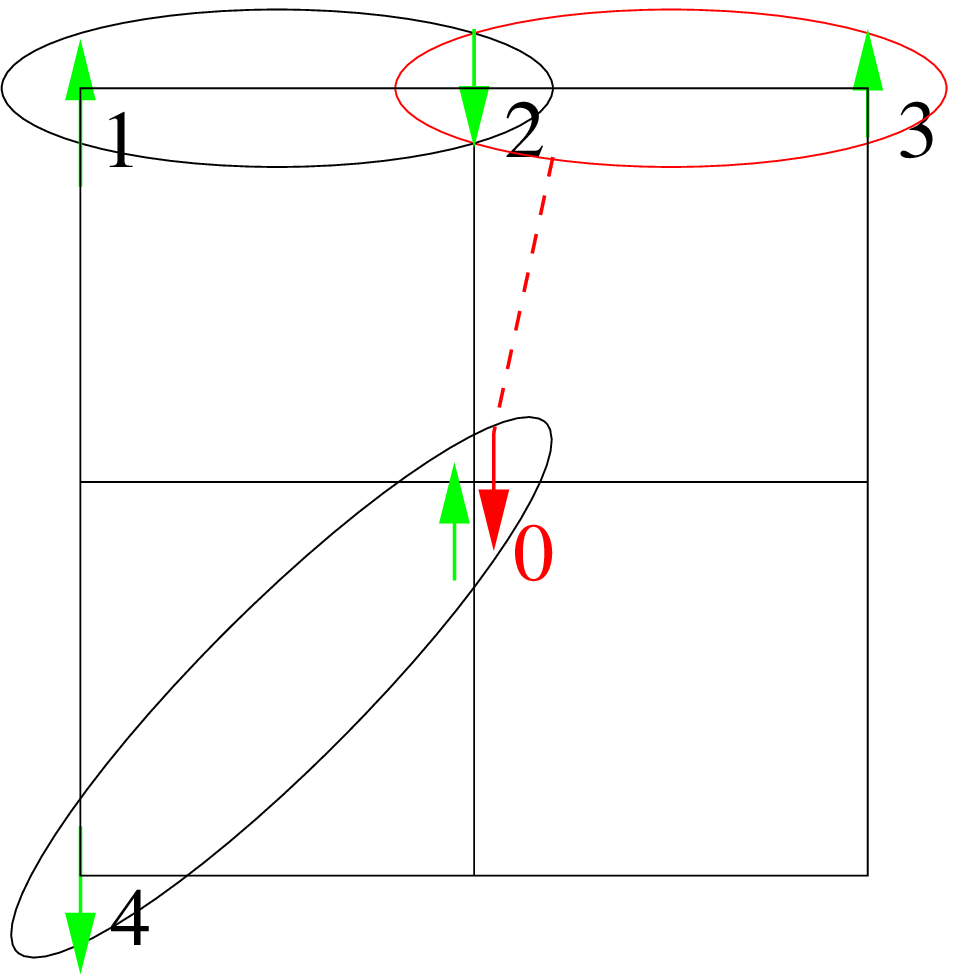} 
\caption{
pSDwf with only $\tilde{\beta}_1$. The site $0$ is empty. $f$-fermions are
represented by green spin, and $d$-fermion is represented by red spin. Black
valence bonds are bonds with of $f$-fermions, while red valence bond has a
$d$-fermion. Notice that the position of $d$-fermion is shifted by $-\hat{y}$
by $\tilde{\beta}_1$ term effect (red dotted line). The two figures show the
two contributions of the overlap between pSDwf and spin basis
$\vert1_{\uparrow}3_{\uparrow}2_{\downarrow}4_{\downarrow}0_{emp}\rangle$, with
\emph{spin of $f$-fermion on site 2 and spin of $d$-fermion on site 0 are
parallel}. They give rise to states: $\mbox{left:
}\vert\uparrow_{1f}\downarrow_{0d}\uparrow_{3f}\downarrow_{2f}\uparrow_{0f}\downarrow_{4f}\rangle=\vert\uparrow_{1f}\uparrow_{3f}\downarrow_{2f}\downarrow_{4f}(\downarrow_{0d}\uparrow_{0f})\rangle$
and $\mbox{right:
}\vert\uparrow_{1f}\downarrow_{2f}\uparrow_{3f}\downarrow_{0d}\uparrow_{0f}\downarrow_{4f}\rangle=-\vert\uparrow_{1f}\uparrow_{3f}\downarrow_{2f}\downarrow_{4f}(\downarrow_{0d}\uparrow_{0f})\rangle
$ The minus sign means the two figures contribute subtractively, i.e., they
cancel exactly.  
} \label{shiftedmixing} 
\end{figure}

\begin{figure}
\includegraphics[width=0.20\textwidth]{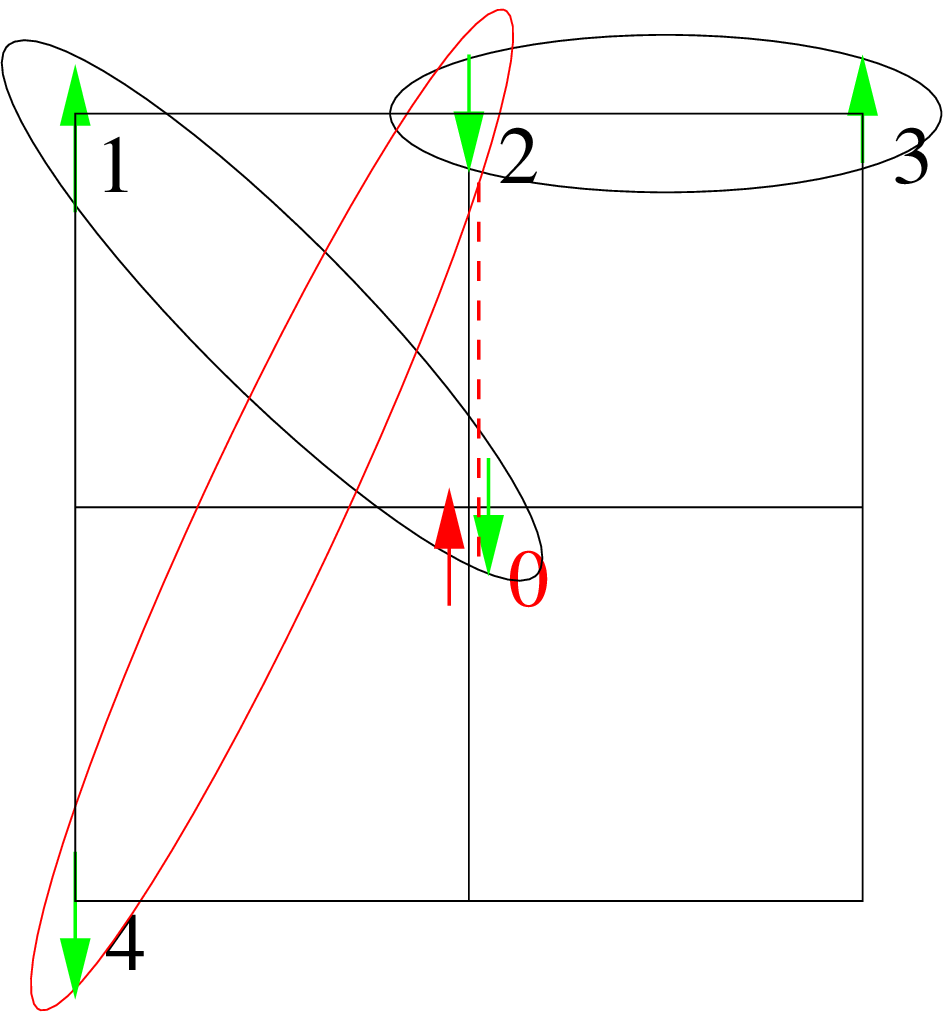} 
\caption{
pSDwf with only $\tilde{\beta}_1$. The site $0$ is empty. $f$-fermions are
represented by green spin, and $d$-fermion is represented by red spin. Black
valence bonds are bonds with of $f$-fermions, while red valence bond has a
$d$-fermion. Note that the position of $d$-fermion is shifted by $-\hat{y}$ by
$\tilde{\beta}_1$ term effect (red dotted line). This figure shows another
contribution to the overlap between pSDwf and spin basis
$\vert1_{\uparrow}3_{\uparrow}2_{\downarrow}4_{\downarrow}0_{emp}\rangle$, with
\emph{spin of $f$-fermion on site 2 and spin of $d$-fermion on site 0 are
anti-parallel.}. It gives rise to state:
$\vert\uparrow_{1f}\downarrow_{0f}\uparrow_{3f}\downarrow_{2f}\uparrow_{0d}\downarrow_{4f}\rangle=-\vert\uparrow_{1f}\uparrow_{3f}\downarrow_{2f}\downarrow_{4f}(\uparrow_{0d}\downarrow_{0f})\rangle$.
Note that there is a contribution for
$\langle1_{\uparrow}3_{\uparrow}0_{\downarrow}4_{\downarrow}2_{emp}\vert\Phi_{PSD}^{SC}(\tilde{\beta}_0)\rangle$,
with the same valence bond map, where
$\vert1_{\uparrow}3_{\uparrow}0_{\downarrow}4_{\downarrow}2_{emp}\rangle$ is
the result state after a hopping along $\hat{y}$ acting on original state
$\vert1_{\uparrow}3_{\uparrow}2_{\downarrow}4_{\downarrow}0_{emp}\rangle$. That
one would give a state
$\vert\uparrow_{1f}\downarrow_{0f}\uparrow_{3f}\downarrow_{2f}\uparrow_{2d}\downarrow_{4f}\rangle=\vert\uparrow_{1f}\uparrow_{3f}\downarrow_{0f}\downarrow_{4f}(\uparrow_{2d}\downarrow_{2f})\rangle$.
The minus sign in the shifted $d$-fermion overlap comes from fermi statistics.
} \label{shiftedmixing_shftok} 
\end{figure}

One immediately sees that the expansion of polynomial Eq.(\ref{pfu}) gives
similar terms as the expansion of Eq.(\ref{pbh}); actually corresponding to one
term in determinant Eq.(\ref{pbd}), we have $2^{N_h}$ terms from
Eq.(\ref{pfu}), since we can either have $\vert\uparrow_f\downarrow_d\rangle$
or $\vert\uparrow_d\downarrow_f\rangle$ for each empty site. The details are
visualized in Fig.\ref{localmixing}. Taking into account the factor$
\frac{1}{\sqrt{2}}$ of projection, one has: 
\begin{widetext} 
\begin{align}
&\langle\{\mathbf{R}_{i\uparrow},\mathbf{R}_{j\downarrow}\}\vert\Phi_{PSD}^{SC}(\tilde{\beta}_0)\rangle
=&(\sqrt{2}\tilde{\beta}_0)^{N_h}\left| 
\begin{array}{cccc}
b(\mathbf{\tilde{R}}_{1\downarrow}-\mathbf{\tilde{R}}_{1\uparrow})&b(\mathbf{\tilde{R}}_{1\downarrow}-\mathbf{\tilde{R}}_{2\uparrow})&\cdots&b(\mathbf{\tilde{R}}_{1\downarrow}-\mathbf{\tilde{R}}_{\frac{N+N_h}{2}\uparrow})\\
b(\mathbf{\tilde{R}}_{2\downarrow}-\mathbf{\tilde{R}}_{1\uparrow})&b(\mathbf{\tilde{R}}_{2\downarrow}-\mathbf{\tilde{R}}_{2\uparrow})&\cdots&b(\mathbf{\tilde{R}}_{2\downarrow}-\mathbf{\tilde{R}}_{\frac{N+N_h}{2}\uparrow})\\
\vdots&\vdots&\ddots&\vdots\\
b(\mathbf{\tilde{R}}_{\frac{N+N_h}{2}\downarrow}-\mathbf{\tilde{R}}_{1\uparrow})&b(\mathbf{\tilde{R}}_{\frac{N+N_h}{2}\downarrow}-\mathbf{\tilde{R}}_{2\uparrow})&\cdots&b(\mathbf{\tilde{R}}_{\frac{N+N_h}{2}\downarrow}-\mathbf{\tilde{R}}_{\frac{N+N_h}{2}\uparrow})
\end{array} \right|,\label{pfd} 
\end{align} 
we found that
$\vert\Phi_{PBCS}\rangle$ and $\vert\Phi_{PSD}^{SC}(\tilde{\beta}_0)\rangle$
are the same wavefunction.

Now let us put in the simplest $k$-dependence in $\tilde{\beta}_k$:
\begin{align} \tilde{\beta}_k=\tilde{\beta}_0+2\tilde{\beta}_1(\cos k_x+\cos
k_y) ,
\end{align} 
we try to write
$\vert\Phi_{PSD}^{SC}(\tilde{\beta}_0,\tilde{\beta}_1)\rangle$ in real space.
After the Fourier transformation into real space: 
\begin{align}
&\vert\Phi_{PSD}^{SC}(\tilde{\beta}_0,\tilde{\beta}_1)\rangle\notag\\
=&P_{SD}\left(\sum_{\mathbf{R}_{i\uparrow},\mathbf{R}_{j\downarrow}}
b(\mathbf{R}_{j\downarrow}-\mathbf{R}_{i\uparrow})(f_{\mathbf{R}_{i\uparrow}\uparrow}^{\dagger}
+\tilde{\beta}_0
d_{\mathbf{R}_{i\uparrow}\uparrow}^{\dagger}+\tilde{\beta}_1\sum_{\mathbf{\delta}=\pm\hat{x},\pm\hat{y}}d_{\mathbf{R}_{i\uparrow+\mathbf{\delta}},\uparrow}^{\dagger}
)(f_{\mathbf{R}_{j\downarrow}\downarrow}^{\dagger}+\tilde{\beta}_0
d_{\mathbf{R}_{j\downarrow}\downarrow}^{\dagger}
+\tilde{\beta}_1\sum_{\mathbf{\delta}=\pm\hat{x},\pm\hat{y}}d_{\mathbf{R}_{j\downarrow+\mathbf{\delta}},\downarrow}^{\dagger})\right)^{\frac{N+N_h}{2}}\vert0\rangle\label{pfb1}.
\end{align} 

If we expand this polynomial Eq.(\ref{pfb1}), of course we will still have
contribution from $\tilde{\beta}_0$ terms which is nothing but the right hand
side of Eq.(\ref{pfd}). But apart from that, we also have contribution from
$\tilde{\beta}_1$, which makes the problem more complicated. To start, let us
consider the case of a single hole $N_h=1$. To have an overlap with spin basis
$\{\mathbf{R}_{i\uparrow},\mathbf{R}_{j\downarrow},\mathbf{R}_{k,0}\}$, the
$d$-fermion on empty site $\mathbf{R}_{k,0}$ can also come from a bond
connecting a spinful site and $\mathbf{R}_{k,0}+\mathbf{\delta}$, which is the
$\tilde{\beta}_1$ term effect. Let us consider the case
$\mathbf{\delta}=\hat{y}$. We can also assume the spin state on site
$\mathbf{R}_{k,0}+\mathbf{\delta}$ is spin down. Now it appears that we
have two ways to construct the empty site on $\mathbf{R}_{k,0}$:
$\vert\uparrow_f\downarrow_d\rangle$ or $\vert\uparrow_d\downarrow_f\rangle$.
We study the two cases separately. Firstly if the empty site is constructed by
$\vert\uparrow_f\downarrow_d\rangle$, shown in Fig.\ref{shiftedmixing}, careful
observation tells us that the contribution to the overlap is exactly cancelled
by fermion statistics.  
On the other hand, if the empty
site is constructed by $\vert\uparrow_d\downarrow_f\rangle$, we have the case
in Fig.\ref{shiftedmixing_shftok}. After careful observation, we know that this
type of contribution is $-\frac{\tilde{\beta}_1}{2\tilde{\beta}_0}$ times the
overlap between $\vert\Phi_{PSD}^{SC}(\tilde{\beta}_0)\rangle$ and the spin
basis that differs from
$\{\mathbf{R}_{i\uparrow},\mathbf{R}_{j\downarrow},\mathbf{R}_{k,0}\}$ by a
hopping along $\hat{y}$. Considering the fact that the shift can also be
$-\hat{y}$ and $\pm\hat{x}$, one has: 
\begin{align}
&\langle\{\mathbf{R}_{i\uparrow},\mathbf{R}_{j\downarrow},\mathbf{R}_{k,0}\}\vert\Phi_{PSD}^{SC}(\tilde{\beta}_0,\tilde{\beta}_1)\rangle\notag\\
=&\sqrt{2}\tilde{\beta}_0\langle\{\mathbf{R}_{i\uparrow},\mathbf{R}_{j\downarrow},\mathbf{R}_{k,0}\}\vert\Phi_{PBCS}\rangle+\left(\frac{-\tilde{\beta}_1}{\sqrt{2}}\right)\langle\{\mathbf{R}_{i\uparrow},\mathbf{R}_{j\downarrow},\mathbf{R}_{k,0}\}\vert\sum_{\mathbf{\delta}=\pm
\hat{x},\pm \hat{y}
}c_{\mathbf{R}_{k,0}+\mathbf{\delta},\alpha}^{\dagger}c_{\mathbf{R}_{k,0},\alpha}^{}\vert\Phi_{PBCS}\rangle\label{pfb1overlap},
\end{align} 
where the minus sign in the second terms comes from Fermi statistics.

Just by looking at Eq.(\ref{pfb1overlap}), we arrive
at the conclusion: 
\begin{align}
\vert\Phi_{PSD}^{SC}(\tilde{\beta}_0,\tilde{\beta}_1)\rangle
=\vert\Phi_{PBCS}\rangle+\left(\frac{-\tilde{\beta}_1}{2\tilde{\beta}_0}
\right)P_D\sum_{i,\mathbf{\delta}
=\pm \hat{x},\pm \hat{y}
}c_{i+\mathbf{\delta},\alpha}^{\dagger}c_{i,\alpha}^{}\vert\Phi_{PBCS}
\rangle.
\label{pfb1overlap1}
\end{align} 

Let us study Eq.(\ref{pfb1overlap}). With out $\tilde{\beta}_1$, one has a
single Slater determinant for the overlap with a spin basis; with
$\tilde{\beta}_1$, we have $1+n_{shift}=5$ Slater determinants, where
$n_{shift}$ is the total number of ways that one hole can hop. Later we will see
that for $n_h$ holes, the number of Slater determinants for the overlap is
$(1+n_{shift})^{n_h}$, which means numerically one can only do few holes.

The result (\ref{pfb1overlap1}) is obtained by studying one-hole case, and
it is not hard to generalize the result for the multi-hole case. Basically,
each hole may either not hop or hop once with a prefactor
$\frac{-\tilde{\beta}_1}{2\tilde{\beta}_0}$, but not hop more than once. For
example, for two-hole case, we have: 
\begin{align}
\vert\Phi_{PSD}^{SC}(\tilde{\beta}_0,\tilde{\beta}_1)\rangle
=\left(1+P_D\sum_{i,\mathbf{\delta}=\pm
\hat{x},\pm \hat{y}
}\frac{-\tilde{\beta}_1}{2\tilde{\beta}_0}c_{i+\mathbf{\delta},\alpha}^{\dagger}c_{i,\alpha}^{}+P_D\frac{1}{2!}\sum_{
\begin{array}{c} 
{\scriptstyle \mathbf{\delta_1},\mathbf{\delta_2}=\pm
\hat{x},\pm \hat{y}}\\ {\scriptstyle j,i\neq j+\mathbf{\delta_2} } 
\end{array}
}\left(\frac{-\tilde{\beta}_1}{2\tilde{\beta}_0}\right)^2
c_{j+\mathbf{\delta_2},\alpha_2}^{\dagger}c_{j,\alpha_2}^{}
c_{i+\mathbf{\delta_1},\alpha_1}^{\dagger}c_{i,\alpha_1}^{}\right)\vert
\Phi_{PBCS}\rangle,
\end{align} 
where the constraint $i\neq j+\mathbf{\delta_2}$
makes sure no hole can hop twice, and the coefficient $\frac{1}{2!}$ comes from
double counting.

We can also easily generalize it to the case with $\tilde{\beta}_2$ and
$\tilde{\beta}_3$.... For two hole case, we have: 
\begin{align}
\vert\Phi_{PSD}^{SC}(\tilde{\beta}_0,\tilde{\beta}_{\mathbf{\delta}})\rangle=\left(1+P_D\sum_{i,\mathbf{\delta}
}\frac{-\tilde{\beta}_{\mathbf{\delta}}}{2\tilde{\beta}_0}c_{i+\mathbf{\delta},\alpha}^{\dagger}c_{i,\alpha}^{}+P_D\frac{1}{2!}\sum_{
\begin{array}{c} {\scriptstyle \mathbf{\delta_1},\mathbf{\delta_2}}\\
{\scriptstyle j,i\neq j+\mathbf{\delta_2} } 
\end{array}
}\left(\frac{-\tilde{\beta}_{\mathbf{\delta_1}}}{2\tilde{\beta}_0}\right)
\left(\frac{-\tilde{\beta}_{\mathbf{\delta_2}}}{2\tilde{\beta}_0}\right)c_{j+\mathbf{\delta_2},\alpha_2}^{\dagger}c_{j,\alpha_2}^{}c_{i+\mathbf{\delta_1},\alpha_1}^{\dagger}c_{i,\alpha_1}^{}\right)\vert\Phi_{PBCS}\rangle.
\end{align} 
In the end, the general formula for multi-hole pSDwf is:
\begin{align}
\vert\Phi_{PSD}^{SC}(\tilde{\beta}_0,\tilde{\beta}_{\mathbf{\delta}})\rangle=P_D
\exp_{n_{hop}=0,1}{\left(1+\sum_{i,\mathbf{\delta}
}\frac{-\tilde{\beta}_{\mathbf{\delta}}}{2\tilde{\beta}_0}c_{i+\mathbf{\delta},\alpha}^{\dagger}c_{i,\alpha}^{}\right)}\vert\Phi_{PBCS}\rangle\label{pfmh},
\end{align} 
where $n_{hop}=0,1$ ensures that no hole can hop twice.
\end{widetext}

\section{How to measure the mixing $\beta_k$? --Physical Meaning of spinon
excitation and dopon excitation} 

From Eq.(\ref{zff+},\ref{zff-}), we know what at the mean-field level,
$\beta_k$ can be measured by spectral weight $Z$. After projection, it is
natural to expect that $Z$ is also closely related to $\beta_k$.  But how to
calculate $Z$ after projection? Basically through Eq.(\ref{zdf+},\ref{zdf-}),
we need a good trial ground state and excited state. The ground state
would be nothing but pSDwf. What is a good excited state? To be specific, let
us study $Z_-$, then the question is how to obtain $\vert N-1\rangle$?

In pBCSwf, the good excited state (referred as quasi-particle state) is found to
be: 
\begin{align} \vert N-1\rangle_{qp}^{PBCS}=P_D c_{p} P_N \vert
BCS\rangle,\label{bcsex} 
\end{align} 
where $P_N$ project into fixed $N_f$ number of electrons. The way we construct
excited state here is simple: first find a excited state on the mean-field level,
then do a projection. In pSDwf, which includes pBCSwf as a limit, we should
have a similar formula. But now we have two possible ways to construct
excitation states, since on mean-field level we have two types of fermions $f$
and $d$, they correspond to two types of excitations. Now it is important to
understand what each type of excitations looks like. It turns out that
\emph{the $f$-type excitation corresponds to the quasi-particle excitation, and
$d$-type excitation corresponds to bare hole excitation}. Thus the
quasi-particle state $\vert N-1\rangle$ for calculating $Z_-$ is the $f$-type
excitation.

The main result for this section is Eq.(\ref{pfex1h},\ref{pfexmh}) for
$f$-excitation and Eq.(\ref{pdex}) for $d$-excitation. One can see that the
$f$-excitation of pSDwf is just the quasi-particle excitation in pBCSwf
together with hopping terms acting on it. And $d$-excitation is the bare hole
on a pSDwf ground state. Let's see how those happen:

For $f$-excitation, 
\begin{align} 
\label{fex}
&\vert N-1\rangle_f=P_{SD} f_{-p}^{\dagger}
P_N \vert \Phi_{SD}^{SC}\rangle 
\\ 
&=P_{SD} f_{-p}^{\dagger} \left(\sum_k
b(k)(f_{k\uparrow}^{\dagger} +\tilde{\beta}_k
d_{k\uparrow}^{\dagger})(f_{-k\downarrow}^{\dagger}+\tilde{\beta}_k
d_{-k\downarrow}^{\dagger})\right)^{\frac{N+N_h}{2}}\vert0\rangle.
\nonumber 
\end{align} 
Here $\vert N-1\rangle_f$ has $N_f-1$ number of electrons, because
before projection there are totally $N+N_h+1$ fermions, and we know projection
enforces one $f$-fermion per site, so totally there are $N_h+1$ $d$-fermions,
i.e., holes. What does this wavefunction look like after projection?

To understand this we first try to understand the excitation of pBCSwf. What is
the excited state in terms of spin basis? One way to see it is to identify:
\begin{align} \vert N-1\rangle_{qp}^{PBCS}&=P_D c_{p\uparrow} \left( \sum_k
a(k)
c_{k\uparrow}^{\dagger}c_{-k\downarrow}^{\dagger}\right)^{N_f/2}\vert0\rangle\\
&\propto P_D c_{-p\downarrow}^{\dagger} \left( \sum_k a(k)
c_{k\uparrow}^{\dagger}c_{-k\downarrow}^{\dagger}\right)^{N_f/2-1}\vert0\rangle.
\end{align} 
In this form the overlap with a spin basis
$\{\mathbf{R}_{i\uparrow},\mathbf{R}_{j\downarrow}\}$ is easy to see. Notice
now the number of up spins is $N_f/2-1$, and the number of down spins is
$N_f/2$, so there is one more site in $\{\mathbf{R}_{j\downarrow}\}$. The only
fashion to construct a spin basis is: let $c_{-p\downarrow}^{\dagger}$ create
an electron somewhere, then let $a(k)
c_{k\uparrow}^{\dagger}c_{-k\downarrow}^{\dagger}$ create the valence bonds.
After observation, the overlap is: 
\begin{widetext}
\begin{align}
\langle\{\mathbf{R}_{i\uparrow},\mathbf{R}_{j\downarrow}\}\vert
N-1\rangle_{qp}^{PBCS}=\left| 
\begin{array}{ccccc}
a(\mathbf{R}_{1\downarrow}-\mathbf{R}_{1\uparrow})&a(\mathbf{R}_{1\downarrow}-\mathbf{R}_{2\uparrow})&\cdots&a(\mathbf{R}_{1\downarrow}-\mathbf{R}_{\frac{N_f}{2}-1,\uparrow})&e^{-ip\mathbf{R}_{1\downarrow}}\\
a(\mathbf{R}_{2\downarrow}-\mathbf{R}_{1\uparrow})&a(\mathbf{R}_{2\downarrow}-\mathbf{R}_{2\uparrow})&\cdots&a(\mathbf{R}_{2\downarrow}-\mathbf{R}_{\frac{N_f}{2}-1,\uparrow})&e^{-ip\mathbf{R}_{2\downarrow}}\\
\vdots&\vdots&\ddots&\vdots\\
a(\mathbf{R}_{\frac{N_f}{2}\downarrow}-\mathbf{R}_{1\uparrow})&a(\mathbf{R}_{\frac{N_f}{2}\downarrow}-\mathbf{R}_{2\uparrow})&\cdots&a(\mathbf{R}_{\frac{N_f}{2}\downarrow}-\mathbf{R}_{\frac{N_f}{2}-1\uparrow})&e^{-ip\mathbf{R}_{\frac{N_f}{2}\downarrow}}
\end{array} \right|.\label{qpdet} 
\end{align}

But to compare with pSDwf formalism, we want to see the same result in a
different way. Let us do a particle-hole transformation, just like what we
did in Eq.(\ref{pbh}).  
\begin{align} 
\vert N-1\rangle_{qp}^{PBCS}=P_D
c_{p\uparrow} \left( \sum_k a(k)
c_{k\uparrow}^{\dagger}c_{-k\downarrow}^{\dagger}\right)^{N_f/2}\vert0\rangle
\propto P_D h_{-p\downarrow}^{\dagger} \left( \sum_k b(k)
h_{k\uparrow}^{\dagger}h_{-k\downarrow}^{\dagger}\right)^{(N+N_h)/2}\vert0
\rangle.
\end{align} 
The only way to construct a spin basis in hole representation
$\{
\mathbf{\tilde{R}}_{i\uparrow},\mathbf{\tilde{R}}_{j\downarrow}\}_h$ is to let
$h_{-p\downarrow}^{\dagger}$ construct a hole somewhere, and let $b(k)
h_{k\uparrow}^{\dagger}h_{-k\downarrow}^{\dagger}$ construct the valence bonds
to fill the lattice. The overlap in hole representation is: 
\begin{align} 
&\langle\{\mathbf{R}_{i\uparrow},\mathbf{R}_{j\downarrow}\}\vert
N-1\rangle_{qp}^{PBCS}=\langle\{\mathbf{\tilde{R}}_{i\uparrow},\mathbf{\tilde{R}}_{j\downarrow}\}_h\vert
N-1\rangle_{qp}^{PBCS} 
\nonumber \\ 
=&\left| 
\begin{array}{ccccc}
b(\mathbf{\tilde{R}}_{1\downarrow}-\mathbf{\tilde{R}}_{1\uparrow})&b(\mathbf{\tilde{R}}_{1\downarrow}-\mathbf{\tilde{R}}_{2\uparrow})&\cdots&b(\mathbf{\tilde{R}}_{1\downarrow}-\mathbf{\tilde{R}}_{\frac{N+N_h}{2}\uparrow})&e^{-ip\mathbf{\tilde{R}}_{1\downarrow}}\\
b(\mathbf{\tilde{R}}_{2\downarrow}-\mathbf{\tilde{R}}_{1\uparrow})&b(\mathbf{\tilde{R}}_{2\downarrow}-\mathbf{\tilde{R}}_{2\uparrow})&\cdots&b(\mathbf{\tilde{R}}_{2\downarrow}-\mathbf{\tilde{R}}_{\frac{N+N_h}{2}\uparrow})&e^{-ip\mathbf{\tilde{R}}_{2\downarrow}}\\
\vdots&\vdots&\ddots&\vdots\\
b(\mathbf{\tilde{R}}_{\frac{N+N_h}{2}+1\downarrow}-\mathbf{\tilde{R}}_{1\uparrow})&b(\mathbf{\tilde{R}}_{\frac{N+N_h}{2}+1,\downarrow}-\mathbf{\tilde{R}}_{2\uparrow})&\cdots&b(\mathbf{\tilde{R}}_{\frac{N+N_h}{2}+1,\downarrow}-\mathbf{\tilde{R}}_{\frac{N+N_h}{2}\uparrow})&e^{-ip\mathbf{\tilde{R}}_{\frac{N+N_h}{2}+1,\downarrow}}
\end{array} \right|.\label{pbexd} 
\end{align} 
\end{widetext}

Now let us go back to pSDwf. What is an $f$-type excitation?  The only way to
construct a spin basis,
$\vert\{\mathbf{\tilde{R}}_{i\uparrow},\mathbf{\tilde{R}}_{j\downarrow}\}_h\rangle$,
is to let $f_{-p}^{\dagger}$ construct an $f$-fermion somewhere, then let
$b(k)(f_{k\uparrow}^{\dagger} +\tilde{\beta}_k
d_{k\uparrow}^{\dagger})(f_{-k\downarrow}^{\dagger}+\tilde{\beta}_k
d_{-k\downarrow}^{\dagger})$ construct the valence bonds to fill the whole
lattice. We first consider the case $\tilde{\beta}_k=\tilde{\beta}_0$. In this
case, the constructing precess is in exactly the same fashion as in
Eq.(\ref{pbexd}), except for one difference: there is a coefficient of
$\sqrt{2}\tilde{\beta}_0$ for each hole. That is because for each hole there
are two contributions, one from
$\vert\uparrow_{f}\downarrow_{d}\rangle$, the other from
$\vert\uparrow_d\downarrow_f\rangle$, each with coefficient of
$\sqrt{1/2}\tilde{\beta}_0$; unless that the hole and the spinon created
by $f_{-p}^{\dagger}$ are on the same site, in which case we have only one
contribution. If we ignore the last effect (since it is an infinitesimal change
to the wavefunction in the low doping limit, and it also comes as an artifact
of our projective construction), we conclude that: 
\begin{widetext}
\begin{align} &\langle\{\mathbf{R}_{i\uparrow},\mathbf{R}_{j\downarrow}\}\vert
N-1,\tilde{\beta}_0\rangle_f=\langle\{\mathbf{\tilde{R}}_{i\uparrow},\mathbf{\tilde{R}}_{j\downarrow}\}_h\vert
N-1,\tilde{\beta}_0\rangle_f\\ =&(\sqrt{2}\beta_0)^{N_h+1}\left|
\begin{array}{ccccc}
b(\mathbf{\tilde{R}}_{1\downarrow}-\mathbf{\tilde{R}}_{1\uparrow})&b(\mathbf{\tilde{R}}_{1\downarrow}-\mathbf{\tilde{R}}_{2\uparrow})&\cdots&b(\mathbf{\tilde{R}}_{1\downarrow}-\mathbf{\tilde{R}}_{\frac{N+N_h}{2}\uparrow})&e^{-ip\mathbf{\tilde{R}}_{1\downarrow}}\\
b(\mathbf{\tilde{R}}_{2\downarrow}-\mathbf{\tilde{R}}_{1\uparrow})&b(\mathbf{\tilde{R}}_{2\downarrow}-\mathbf{\tilde{R}}_{2\uparrow})&\cdots&b(\mathbf{\tilde{R}}_{2\downarrow}-\mathbf{\tilde{R}}_{\frac{N+N_h}{2}\uparrow})&e^{-ip\mathbf{\tilde{R}}_{2\downarrow}}\\
\vdots&\vdots&\ddots&\vdots\\
b(\mathbf{\tilde{R}}_{\frac{N+N_h}{2}+1\downarrow}-\mathbf{\tilde{R}}_{1\uparrow})&b(\mathbf{\tilde{R}}_{\frac{N+N_h}{2}+1,\downarrow}-\mathbf{\tilde{R}}_{2\uparrow})&\cdots&b(\mathbf{\tilde{R}}_{\frac{N+N_h}{2}+1,\downarrow}-\mathbf{\tilde{R}}_{\frac{N+N_h}{2}\uparrow})&e^{-ip\mathbf{\tilde{R}}_{\frac{N+N_h}{2}+1,\downarrow}}
\end{array} \right|\\
&=(\sqrt{2}\beta_0)^{N_h+1}\langle\{\mathbf{R}_{i\uparrow},\mathbf{R}_{j\downarrow}\}\vert
N-1\rangle_{qp}^{PBCS},\label{pfexf} 
\end{align} 
so 
\begin{align}
\vert N-1,\tilde{\beta}_0\rangle_f\propto \vert N-1\rangle_{qp}^{PBCS}.
\end{align} The point is that $f$-type excitation describes the spin-charge
separation picture of the excitation, because the hole and the unpaired spinon
created by $f_{-p}^{\dagger}$ can be arbitrarily separated. And it turns out to
be the low energy excitation of $t$-$J$ model.

What if the mixing $\beta_k$ has momentum dependence? Similar to our study for
the ground state wavefunction leading to
Eq.(\ref{pfb1overlap},\ref{pfb1overlap1}), one can convince oneself that, in
the one hole case 
\begin{align} 
&\vert
N-1,\tilde{\beta}_0,\tilde{\beta}_1,\tilde{\beta}_2,\tilde{\beta}_3\rangle_f\propto
\vert
N-1\rangle_{qp}^{PBCS}+\left(\frac{-\tilde{\beta}_1}{2\tilde{\beta}_0}\right)P_D\sum_{i,\mathbf{\delta}=\pm
\hat{x},\pm \hat{y} }c_{i+\mathbf{\delta},\alpha}^{\dagger}c_{i,\alpha}^{}\vert
N-1\rangle_{qp}^{PBCS}\notag\\
&+\left(\frac{-\tilde{\beta}_2}{2\tilde{\beta}_0}\right)P_D\sum_{i,\mathbf{\delta}=\pm
(\hat{x}+\hat{y}),\pm (\hat{x}-\hat{y})
}c_{i+\mathbf{\delta},\alpha}^{\dagger}c_{i,\alpha}^{}\vert
N-1\rangle_{qp}^{PBCS}+\left(\frac{-\tilde{\beta}_3}{2\tilde{\beta}_0}\right)P_D\sum_{i,\mathbf{\delta}=\pm
2\hat{x},\pm 2\hat{y}
}c_{i+\mathbf{\delta},\alpha}^{\dagger}c_{i,\alpha}^{}\vert
N-1\rangle_{qp}^{PBCS}.\label{pfex1h} 
\end{align} 

And for multi-hole case,
similar to Eq.(\ref{pfmh}) 
\begin{align} \vert
N-1,\tilde{\beta}_0,\tilde{\beta}_{\delta}\rangle_f\propto P_D
\exp_{n_{hop}=0,1}{\left(1+\sum_{i,\mathbf{\delta}
}\frac{-\tilde{\beta}_{\mathbf{\delta}}}{2\tilde{\beta}_0}c_{i+\mathbf{\delta},\alpha}^{\dagger}c_{i,\alpha}^{}\right)}\vert
N-1\rangle_{qp}^{PBCS}.\label{pfexmh} 
\end{align}

For $d$-excitation, story is different. It turns out $d$-excitation corresponds
to bare hole excitation. What is a bare hole excitation $\vert
N-1\rangle_{bh}$? For a pBCSwf, 
\begin{align} \vert
N-1\rangle_{bh}^{PBCS}&=c_{p\uparrow}\vert\Phi_{PBCS}\rangle ,
\end{align} in
terms of spin basis, it is easy to show that: 
\begin{align}
&\langle\{\mathbf{R}_{i\uparrow},\mathbf{R}_{j\downarrow},\mathbf{R}_{k0}\}\vert
N-1\rangle_{bh}^{PBCS}=\langle\{\mathbf{R}_{i\uparrow},\mathbf{R}_{j\downarrow},\mathbf{R}_{k0}\}\vert
c_{p\uparrow} \vert\Phi_{PBCS} \rangle
\propto \sum_k
e^{-ip\mathbf{R}_{k0}}\langle\{\{\mathbf{R}_{i\uparrow},\mathbf{R}_{k0}\},\mathbf{R}_{j\downarrow}\}\vert\Phi_{PBCS}
\rangle
\nonumber \\ 
=&\sum_ke^{-ip\mathbf{R}_{k0}}\left| 
\begin{array}{ccccc}
a(\mathbf{R}_{1\downarrow}-\mathbf{R}_{1\uparrow})&a(\mathbf{R}_{1\downarrow}-\mathbf{R}_{2\uparrow})&\cdots&a(\mathbf{R}_{1\downarrow}-\mathbf{R}_{\frac{N_f}{2}-1\uparrow})&a(\mathbf{R}_{1\downarrow}-\mathbf{R}_{k0})\\
a(\mathbf{R}_{2\downarrow}-\mathbf{R}_{1\uparrow})&a(\mathbf{R}_{2\downarrow}-\mathbf{R}_{2\uparrow})&\cdots&a(\mathbf{R}_{2\downarrow}-\mathbf{R}_{\frac{N_f}{2}-1\uparrow})&a(\mathbf{R}_{2\downarrow}-\mathbf{R}_{k0})\\
\vdots&\vdots&\ddots&\vdots&\vdots\\
a(\mathbf{R}_{\frac{N_f}{2}\downarrow}-\mathbf{R}_{1\uparrow})&a(\mathbf{R}_{\frac{N_f}{2}\downarrow}-\mathbf{R}_{2\uparrow})&\cdots&a(\mathbf{R}_{\frac{N_f}{2}\downarrow}-\mathbf{R}_{\frac{N_f}{2}-1\uparrow})&a(\mathbf{R}_{\frac{N_f}{2}\downarrow}-\mathbf{R}_{k0})
\end{array} \right|.\label{bhdet} 
\end{align}

What is a $d$-type excitation in terms of spin basis? We first consider the
case where $\tilde{\beta}_k=\tilde{\beta}_0$, 
\begin{align} 
\vert N-1,\tilde{\beta}_0\rangle_d=P_{SD} d_{-p\downarrow}^{\dagger} P_N \vert
\Phi_{SD}^{SC}\rangle 
=P_{SD} d_{-p\downarrow}^{\dagger} \left(\sum_k
b(k)(f_{k\uparrow}^{\dagger} +\tilde{\beta}_k
d_{k\uparrow}^{\dagger})(f_{-k\downarrow}^{\dagger}+\tilde{\beta}_k
d_{-k\downarrow}^{\dagger})\right)^{\frac{N+N_h}{2}}\vert0\rangle.\label{dex}
\end{align} 
\end{widetext}
One can see that the only way to construct a spin basis,
$\vert\{\mathbf{\tilde{R}}_{i\uparrow},\mathbf{\tilde{R}}_{j\downarrow}\}_h\rangle$,
is to let $b(k)(f_{k\uparrow}^{\dagger} +\tilde{\beta}_k
d_{k\uparrow}^{\dagger})(f_{-k\downarrow}^{\dagger}+\tilde{\beta}_k
d_{-k\downarrow}^{\dagger})$ construct the valence bonds to fill the whole
lattice, then find a site occupied by one $f_{\uparrow}$-fermion only, and let
$d_{-p\downarrow}^{\dagger}$ construct a hole there. By observation, we
conclude, 
\begin{align} 
\vert N-1,\tilde{\beta}_0\rangle_d \propto \vert N-1\rangle_{bh}^{PBCS}.
\end{align}

If $\tilde{\beta}_k$ has $k$-dependence, one can also convince oneself that
\begin{align} \vert N-1,\tilde{\beta}_0,\tilde{\beta}_{\delta}\rangle_d\propto
c_p\vert\Phi_{PSD}^{SC}(\tilde{\beta}_0,\tilde{\beta}_{\mathbf{\delta}})\rangle\equiv\vert
N-1\rangle_{bh}^{PSD}\label{pdex},
\end{align} i.e., $d$-type excitation
corresponds to the bare hole in pSDwf.

To summarize, we have the following identification: $f$-type excitation
corresponds to the low energy quasi-particle excitation, i.e., a state
constructed by putting $c_p$ operator inside the projection; $d$-type
excitation corresponds to the bare hole excitation, i.e., a state constructed
by putting $c_p$ operator outside the projection.

\section{Numerical Methods and Results}\label{NM} We use Variational Monte
Carlo (VMC) method to calculate the ground state energy (of 2 holes), the
excited state energy (of 1 hole) of pSDwf and pBCSwf and the spectral weight
$Z_{-,k}$. 

Our pBCSwf calculation is mostly traditional. Nevertheless the previous
calculation of $Z_{-}$\cite{bieri-2006} is indirect and having uncontrolled
error bars inside the fermi surface. We developed a straightforward technique
to calculate $Z_{-}$. Let us recall the definition of $Z_{-}$ Eq.(\ref{zdf-}).
For pBCSwf, if we relabel $\vert N-1\rangle_{bh}^{PBCS}$ as $\vert bh\rangle$
and $\vert N-1\rangle_{qp}^{PBCS}$ as $\vert qp\rangle$ to save notation:
\begin{align} Z_{-,k}&=\frac{\vert\langle qp\vert bh\rangle\vert^2}{\langle
qp\vert qp\rangle\langle \Phi_{PBCS}\vert\Phi_{PBCS}\rangle}\notag\\
&=\frac{\vert\langle qp\vert bh\rangle\vert^2}{\langle qp\vert qp\rangle\langle
bh\vert bh\rangle}\frac{\langle bh\vert bh\rangle}{\langle
\Phi_{PBCS}\vert\Phi_{PBCS}\rangle}\notag\\ &=\frac{\vert\langle qp\vert
bh\rangle\vert^2}{\langle qp\vert qp\rangle\langle bh\vert bh\rangle} n_k,
\end{align} where $n_k$ is the occupation number of particles at momentum $k$.
$n_k$ can be calculated by VMC approach pretty
straightforwardly\cite{01536790593}. In particular, one can easily show that at
low doping limit, which is the case considered in this paper, $n_k=\frac{1}{2}$
independent of $k$ exactly. The only thing one needs to worry about is the
overlap prefactor between $\vert qp\rangle$ and $\vert bh\rangle$. Instead of
calculating the factor itself, one can split the calculation
into two. If we denote a spin basis as $\vert s\rangle$, 
\begin{widetext}
\begin{align}
\frac{\langle qp\vert bh\rangle}{\langle qp\vert
qp\rangle}=\sum_{s}\frac{\langle qp\vert s\rangle\langle s\vert
bh\rangle}{\langle qp\vert qp\rangle}=\sum_{s}\frac{\langle s\vert
bh\rangle}{\langle s\vert qp\rangle}\frac{\vert\langle qp\vert
s\rangle\vert^2}{\langle qp\vert qp\rangle}=\sum_{s}\frac{\langle s\vert
bh\rangle}{\langle s\vert qp\rangle}\rho_{qp}(s),\label{fdoverff}\\
\frac{\langle bh\vert qp\rangle}{\langle bh\vert
bh\rangle}=\sum_{s}\frac{\langle bh\vert s\rangle\langle s\vert
qp\rangle}{\langle bh\vert bh\rangle}=\sum_{s}\frac{\langle s\vert
qp\rangle}{\langle s\vert bh\rangle}\frac{\vert\langle bh\vert
s\rangle\vert^2}{\langle bh\vert bh\rangle}=\sum_{s}\frac{\langle s\vert
qp\rangle}{\langle s\vert bh\rangle}\rho_{bh}(s).\label{fdoverdd} 
\end{align}
\end{widetext}
Since both $\langle s\vert qp\rangle$ and $\langle s\vert bh\rangle$ are Slater
determinant or sum of Slater determinants (see Eq.(\ref{qpdet}) and
Eq.(\ref{bhdet})), the above two quantities can be calculated by Metropolis
program in a straightforward fashion. Then the product of the two gives the
$Z_{-,k}$. This algorithm works for finite doping case, too.

For pSDwf, because we include $k$-dependent mixing, in each step of Metropolis
random walk, we need to keep track of all the $(1+n_{shift})^{n_{h}}$ matrices,
which limit the calculation for few holes.  \subsection{Ground state at half
filling and 2 holes} The calculation is done for $t$-$t'$-$t''$-$J$ model on 10
by 10 lattice, where $t=1$, $t'=-0.3$, $t''=0.2$ and $J=0.3$. We choose
periodic boundary condition in x-direction, and anti-periodic boundary
condition in y-direction. 

\begin{table*} 
\begin{tabular}{|l|l|} \hline energy per bond&$\vec S_i\cdot\vec
S_{i+1}$ \\ \hline -0.1710$\pm$0.0001&-0.3200$\pm$0.0001\\ \hline 
\end{tabular}
\caption{half-filling ground state on 10 by 10 lattice} \label{hftb}
\end{table*}

For variational parameters, we choose the lowest-energy ansatz in
Eq.(\ref{dwave})\cite{ivanov:132501} with parameters 
\begin{align}
&\chi=1,&&\Delta=0.55,&&\mu=0.  
\end{align} The energy for half-filling ground
state is listed in Table \ref{hftb}.

\begin{table*} 
\begin{tabular}{|l|l|l|l|l|l|l|l|l|l|l|l|} \hline
wavefunction&$\frac{\Delta}{\chi}$&$\frac{\chi'}{\chi}$&$\frac{\chi''}{\chi}$&$\frac{\tilde{\beta}_1}{\tilde{\beta}_0}$&$\frac{\tilde{\beta}_2}{\tilde{\beta}_0}$&$\frac{\tilde{\beta}_3}{\tilde{\beta}_0}$&$\begin{array}{c}\mbox{total energy}\\
\mbox{per bond}\end{array}$&$\begin{array}{c}\la\vec S_i\cdot\vec S_{i+1}\ra\\\mbox{per bond}\end{array}$&$T_1$&$T_2$&$T_3$ \\ \hline
pBCSwf&0.55&0&0&0&0&0&-0.1872$\pm$0.0001&   -0.2977$\pm$0.0002 &
2.64$\pm$0.01 &   0.52$\pm$0.01  &  0.48$\pm$0.01\\ \hline
pBCSwf(optimal)&0.55&-0.4&0.0&0&0&0&-0.1890$\pm$0.0001 &  -0.2947$\pm$0.0002 &
2.66$\pm$0.01   &0.06$\pm$0.01 &    1.07$\pm$0.01\\\hline
pBCSwf&0.55&-0.5&0.1&0&0&0&-0.1885$\pm$0.0001 &  -0.2872$\pm$0.0002 &
2.66$\pm$0.01   &-0.23$\pm$0.01 &    1.52$\pm$0.01\\\hline
pSDwf(optimal)&0.55&0&0&-0.3 & 0.3 & -0.1 & -0.1918$\pm$0.0001 &
-0.2943$\pm$0.0002 &      2.86$\pm$0.01& -0.46$\pm$0.01&
0.77$\pm$0.01\\\hline
\end{tabular} 
\caption{ 
Two holes on 10 by 10 lattice. $t=1$, $t'=-0.3$, $t''=0.2$ and $J=0.3$. $T_1$, $T_2$ and $T_3$ stand for nearest neighbor hopping per hole $\frac{1}{N_h}\sum_{i,\delta=\pm\hat{x},\pm\hat{y}}\la c_i^{\dag}c_{i+\delta}\ra$, next nearest neighbor hopping per hole $\frac{1}{N_h}\sum_{i,\delta=\pm(\hat{x}+\hat{y}),\pm(\hat{x}-\hat{y})}\la c_i^{\dag}c_{i+\delta}\ra$ and third nearest neighbor hopping per hole $\frac{1}{N_h}\sum_{i,\delta=\pm2\hat{x},\pm2\hat{y}}\la c_i^{\dag}c_{i+\delta}\ra$ respectively. We
compare pBCSwf of $d$-wave ansatz, pBCSwf with longer range hoppings $\chi'$ and
$\chi''$, and pSDwf with non-local mixings. The best trial pSDwf has an energy
1.5\% below that of the best trial pBCSwf with longer range hoppings.  Comparing the
first and the last line which have the same spin correlation, we find that the
energy of a hole in pSDwf is $0.46t$ lower than that of a hole in pBCSwf.  Note
that pBCSwf with longer range hoppings destroys the $d$-wave spin background. As a
result, the attempt to lower the hopping energy by tuning $\chi'$ and $\chi''$
is not effective since it would increase the spin energy a lot.  
} 
\label{thtb}
\end{table*}

For two holes, we compare the energy of ground states of pBCSwf and pSDwf. For
pSDwf, to lower the $t$ hopping energy, since $t<0$, by Eq.(\ref{pfmh}), the
sign of $\tilde{\beta}_1$ should be negative. Similarly since $t'>0, t''<0$,
the parameters lowering $t'$ and $t''$ hopping energy have signs
$\tilde{\beta}_2>0$ and $\tilde{\beta}_3<0$. We did a variational search for
the optimal values of $\tilde{\beta}_i$. The results are listed in Table
\ref{thtb}, where we also compare it with pBCSwf with longer range hoppings (see
Section \ref{diff}).

We find that the energy of the best pSDwf is lower than the energy of the best
pBCSwf.  We note that the pSDwf and pBCSwf are identical at half filling.  So
the energy difference between the two states is purely a doping effect.
Comparing the first and the last line in table \ref{thtb} which have the same
spin correlation, we see that the total energies of the two states differ by
$0.0046\times 200$ since the the 10 by 10 lattice has 200 links.
This energy difference is due to the presence of two holes.
So the energy of a hole in pSDwf
is $0.46t$ lower than that of a hole
in pBCSwf. This energy difference is big, indicating that the
charge-spin correlation is much better described by pSDwf than pBCSwf.


\subsection{Hole doped case, quasi-particle excitations and $Z_-$.}

In this section we study the excitations of $t$-$t'$-$t''$-$J$ model, which is
one hole on 10x10 lattice. We also compare pSDwf with pBCSwf. We know from
Eq.(\ref{pfexmh}) that the pSDwf $f$-excitation state goes back to pBCSwf
quasi-particle excitation state when all non-local mixings
$\tilde{\beta}_{\delta}=0$. Also from Eq.(\ref{pfexmh}), one can see that to
lower the $t$, $t'$ and $t''$ hopping energy, we should also have
$\tilde{\beta}_1<0$, $\tilde{\beta}_2>0$ and $\tilde{\beta}_3<0$. Actually in
the low doping limit, one should expect the non-local mixing
$\tilde{\beta}_{\delta}$ for quasi-particle excited states ($f$-excitation) to
be same as the ground state. Here we adopt the values of
$\tilde{\beta}_{\delta}$ from our study of 2-hole system ground state. 

Our VMC calculation shows that the pSDwf or pBCSwf has finite $Z_{-}$ deep
inside the fermi surface even in the low doping limit $x\rightarrow 0$. This is
physically wrong because deep inside fermi surface there is no well-defined
quasi-particle, and the idea of calculating $Z_-$ by a single particle excited
state is also incorrect. Nevertheless, because the low energy excitation is
more and more quasi-particle like as one approaches the fermi surface, we
expect that the $Z_{-}$ calculation remains valid close to fermi surface,
roughly speaking, along the diagonal direction from $(\pi,0)$ to $(0,\pi)$.

\begin{figure}
\begin{center}
\includegraphics[width=0.3\textwidth]{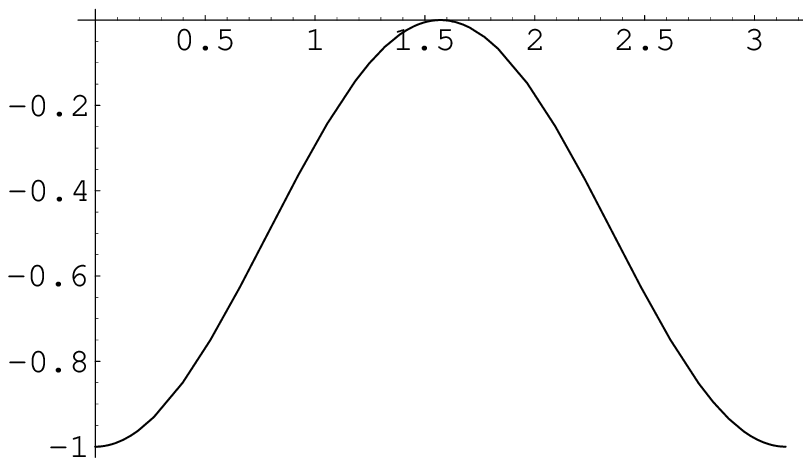}\\
\hskip 5mm \includegraphics[width=0.3\textwidth]{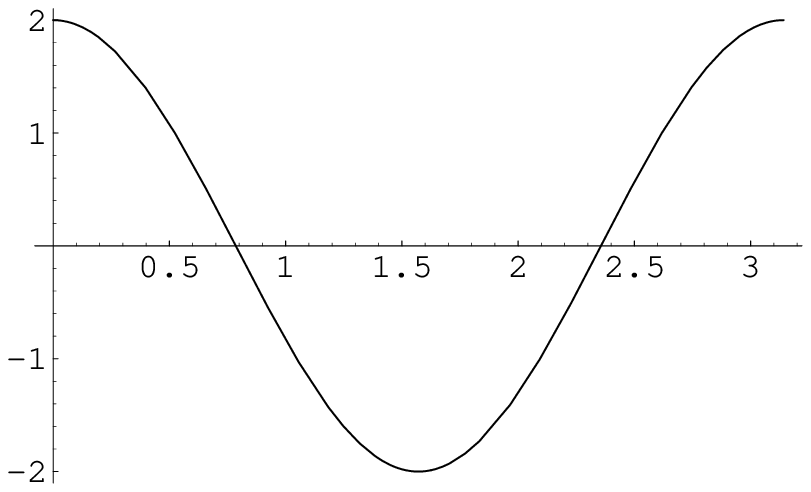}
\end{center}
\caption{Shapes of functions $\cos k_x\cos k_y$ (up) and $\cos 2k_x+\cos
2k_y$ (down) along diagonal direction from $(\pi,0)$ to $(0,\pi)$.}
\label{cosplot} 
\end{figure}

From Eq.(\ref{zff-}), we know that at the mean-field level, the modulation of
$Z_-$ is controlled by $\beta_k^2$. It is important to study the shapes of
$\beta_k^2$ for various cases. In Fig. \ref{cosplot} we plot the shapes of
functions $\cos k_x\cos k_y$ and $\cos 2k_x+\cos 2k_y$ along the diagonal
direction. If $\beta_k=\beta_0+2\beta_1(\cos k_x+\cos k_y)$, $\beta_k^2$
remains constant along the diagonal direction. If $\beta_k=\beta_0+4\beta_2\cos
k_x\cos k_y$, for small $\beta_2>0$, $\beta_k^2$ is reduced at the anti-nodal
point. If $\beta_k=\beta_0+2\beta_3(\cos 2 k_x+\cos 2 k_y)$, for small
$\beta_3<0$, $\beta_k^2$ is enhanced at the nodal point and suppressed at the
anti-nodal point. Let us remember this trend: positive $\beta_2$ and negative
$\beta_3$ drive the modulation of $Z_-$ in the way consistent with dichotomy
for hole doped samples.

For small values of $\beta_k$ we know that $\tilde{\beta}_k\approx\beta_k$.
Eq.(\ref{zff-}) suggests that along the diagonal direction 
\begin{align}
Z_{-,k}\propto (\tilde{\beta}_0+4\tilde{\beta}_2\cos k_x\cos
k_y+2\tilde{\beta}_3(\cos 2 k_x+\cos 2 k_y))^2 .
\end{align} But as a mean-field
result, one should expect that the above equation is only valid qualitatively. In
fact to crudely fit the relation of the modulation of $Z_-$ and
$\tilde{\beta}$, we found it is better to have some order of unity extra factor
in front of $\tilde{\beta}$ terms, and $\tilde{\beta}_1$ also contributes to
the modulation of $Z_-$ as a uniform shift.  
\begin{align} Z_{-,k}\propto
(\tilde{\beta}_0+\frac{\tilde{\beta}_1}{2}+\tilde{\beta}_2\cos k_x\cos
k_y+\frac{\tilde{\beta}_3}{2}(\cos 2 k_x+\cos 2 k_y))^2.  
\end{align}

For $t$-$J$ model without $t'$ and $t''$, there is no reason to develop a
finite value of $\beta_2$ and $\beta_3$ since there is no longer range hoppings. As
a result, one expects that $Z_-$ remains almost constant along the diagonal
direction.

For $t$-$t'$-$t''$-$J$ model with $t>0$, $t'<0$, $t''>0$, we know that
$\tilde{\beta}_2>0$ and $\tilde{\beta}_3<0$ have to be developed to favor
longer range hoppings. So one expect $Z_-$ should develop dichotomy shape along the
diagonal direction.

\begin{figure}
\includegraphics[width=0.22\textwidth]{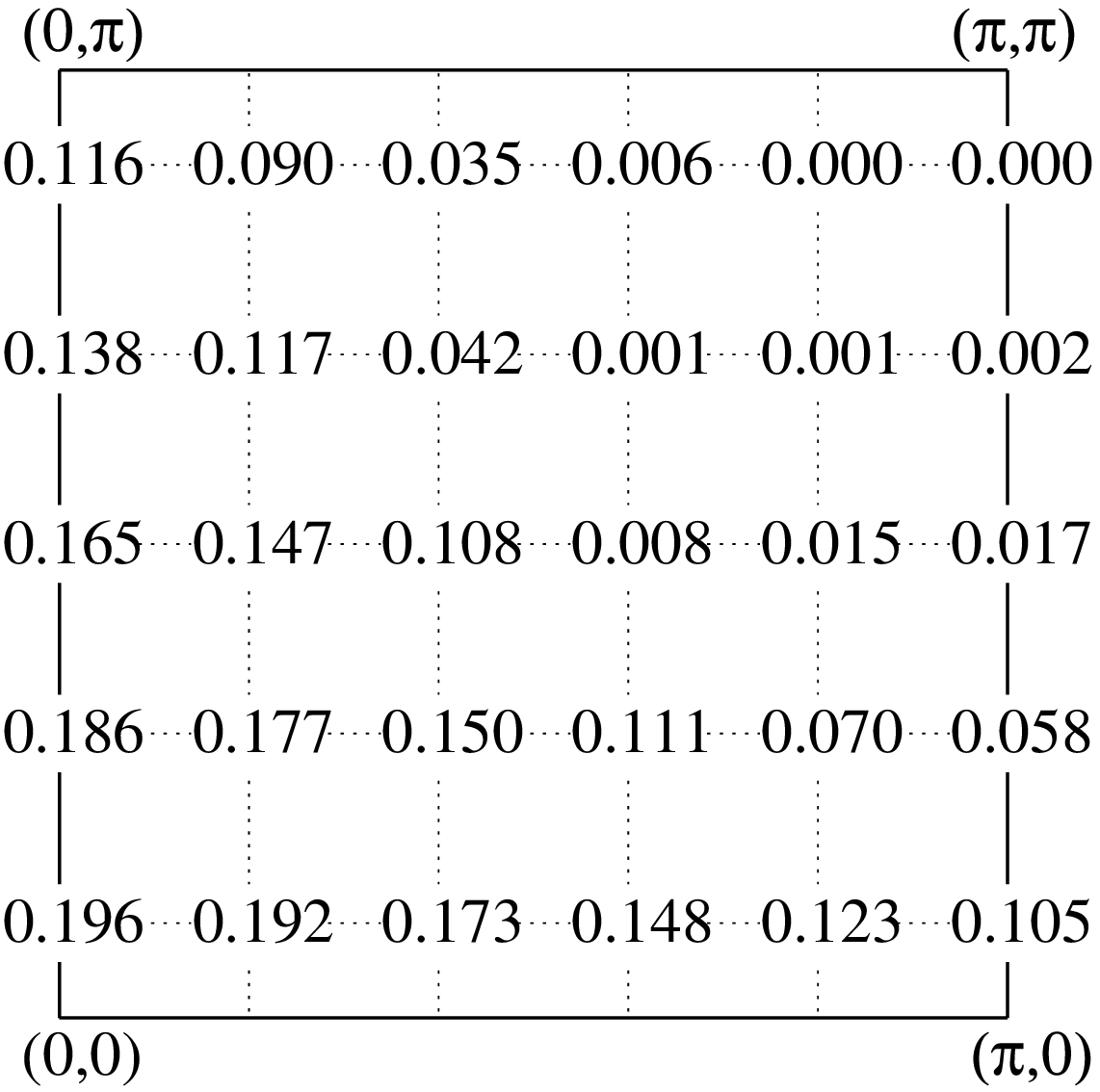}\;\;\;\;\;
\includegraphics[width=0.22\textwidth]{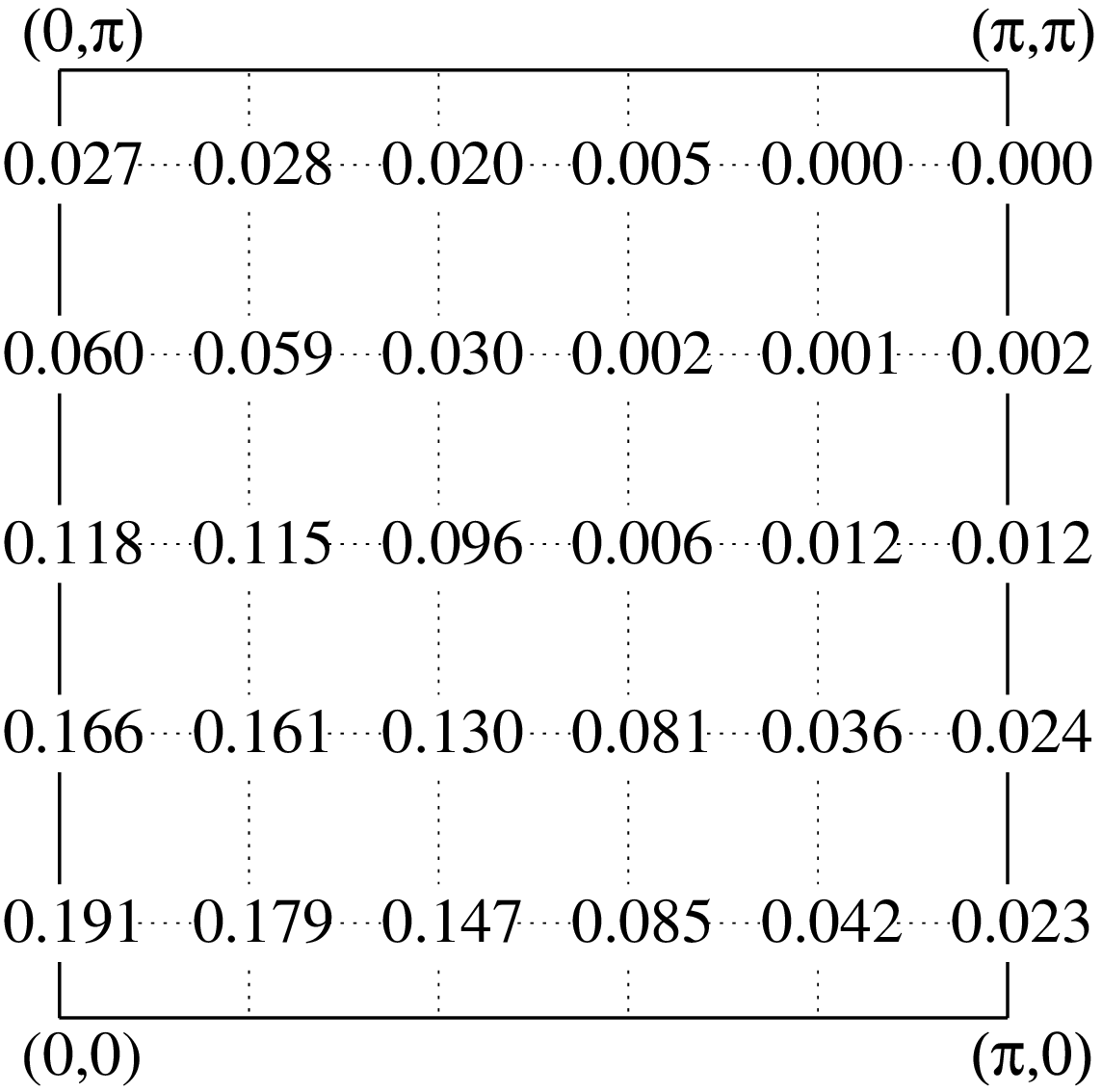}
\caption{For one hole on 10 by 10 lattice, we plot $Z_{-,k}$ of pBCSwf (left,
$\chi=1$, $\Delta=0.55$) and pSDwf (right, $\chi=1$, $\Delta=0.55$,
$\frac{\tilde{\beta}_1}{\tilde{\beta}_0}=-0.3$,
$\frac{\tilde{\beta}_2}{\tilde{\beta}_0}=0.3$,
$\frac{\tilde{\beta}_1}{\tilde{\beta}_0}=-0.1$). pBCSwf has almost constant
$Z_-$ along diagonal direction from $(\pi,0)$ to $(0,\pi)$; while pSDwf has
$Z_-$ suppressed at anti-nodal point.} \label{z-plot} 
\end{figure}

In Fig.\ref{z-plot} we compare the $Z_{-,k}$ of pBCSwf and pSDwf. One can see
that pSDwf shows strong dichotomy.

\begin{figure*}
\includegraphics[height=2.2in]{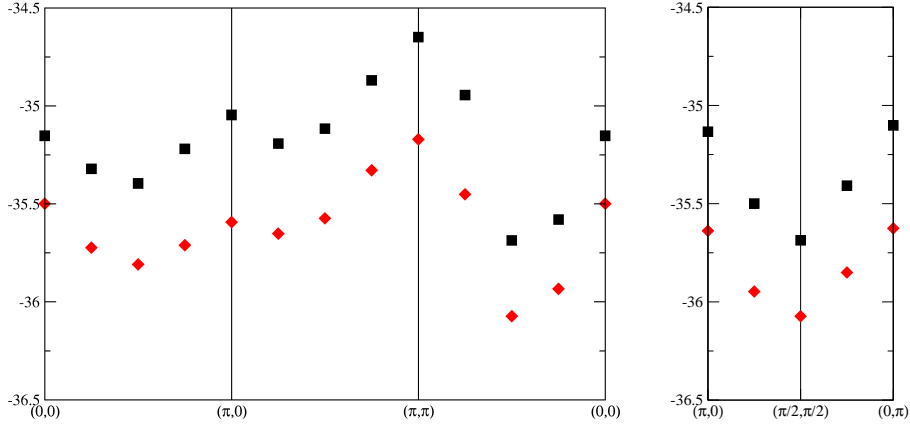}\;\;\;
\includegraphics[height=2.2in]{excitation2.eps}
\caption{
Quasi-particle spectrum for one hole on 10 by 10 lattice. $t=1$, $t'=-0.3$,
$t''=0.2$ and $J=0.3$. The black square shows the spectrum of $d$-wave pBCSwf
with $\chi=1$ and $\Delta=0.55$, and the red diamond shows the spectrum of
pSDwf with $\chi=1$, $\Delta=0.55$,
$\frac{\tilde{\beta}_1}{\tilde{\beta}_0}=-0.3$,
$\frac{\tilde{\beta}_2}{\tilde{\beta}_0}=0.3$ and
$\frac{\tilde{\beta}_3}{\tilde{\beta}_0}=-0.1$. One can see the first hole
doped to $(\pi/2,\pi/2)$.  The energy of a doped hole in pSDwf is $0.38t$ lower
than that of a hole in pBCSwf.
} \label{excitation} 
\end{figure*}

In Fig.\ref{excitation} we compare the energy dispersion of one-hole
quasi-particle excitations of pBCSwf and pSDwf. 
The energy of a doped hole in pSDwf is $0.38t$ lower
than that of a hole in pBCSwf.

\subsection{Electron doped case} 

In electron-doped case, one can do a
particle-hole transformation, then multiply a $(-1)$ for the odd lattice
electron operators. By doing so, the original electron-doped $t$-$J$ model with
parameters $t,t',t'',J$ transformed into hole-doped $t$-$J$ model with
parameters $t,-t',-t'',J$, together with a $(\pi,\pi)$ shift in momentum space. 

The approach outlined in Eq.(\ref{fdoverff}) and Eq.(\ref{fdoverdd}) still
applies here. But because of the particle-hole transformation, we are
calculating $Z_{+}$ of the original electron-doped system. Because $t'>0$ and
$t''<0$, to favor longer range hoppings, we must have $\tilde{\beta}_2<0$ and
$\tilde{\beta}_3>0$, which differ from hole-doped case by a sign flip. As a
result, the $Z_{+}$ now will be suppressed at nodal point, but enhanced at the
anti-nodal point. This is exactly what people observed in exact
diagonalization\cite{PhysRevB.56.6320}.

\begin{figure}
\includegraphics[width=0.22\textwidth]{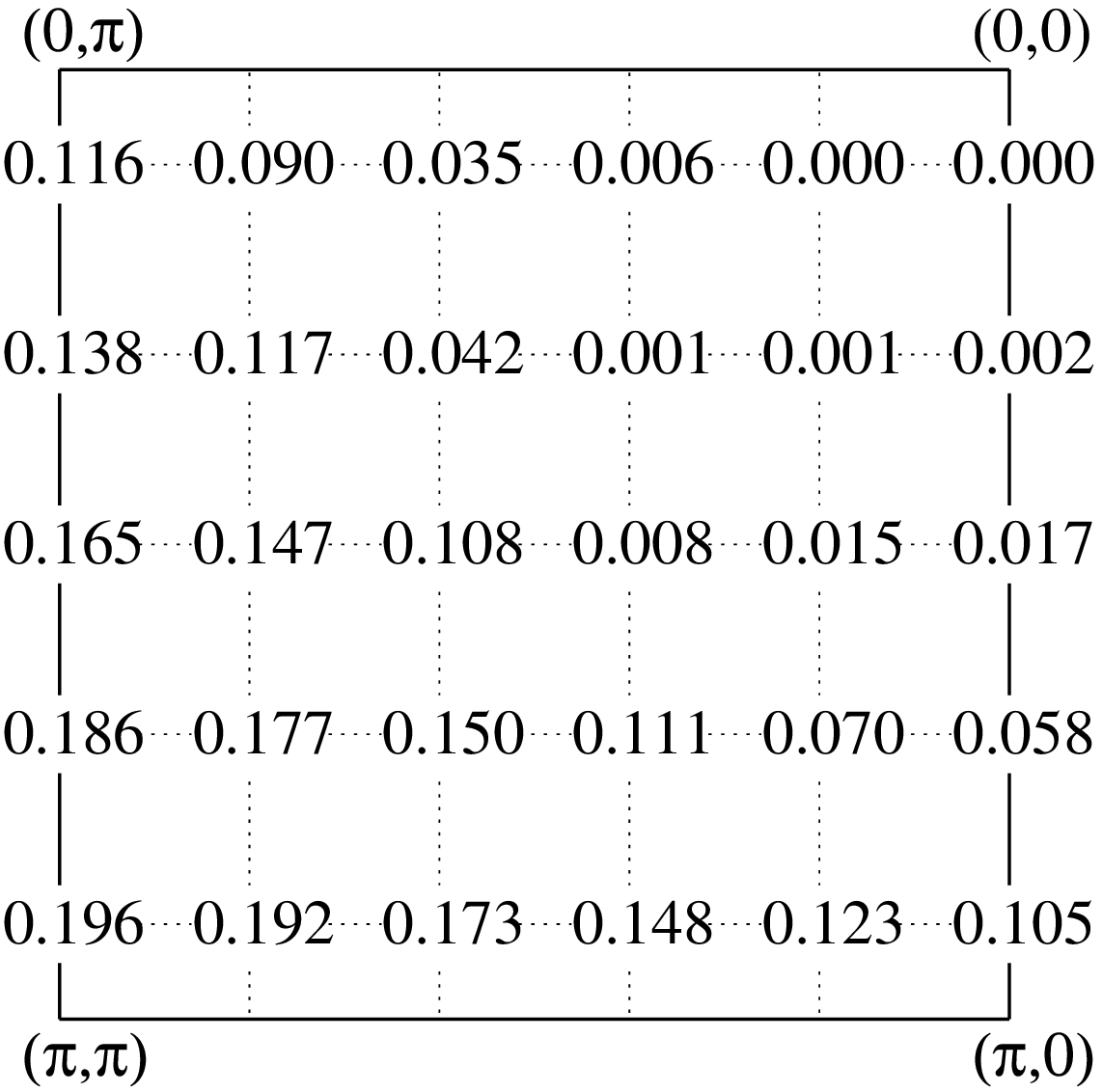}\;\;\;\;\;
\includegraphics[width=0.22\textwidth]{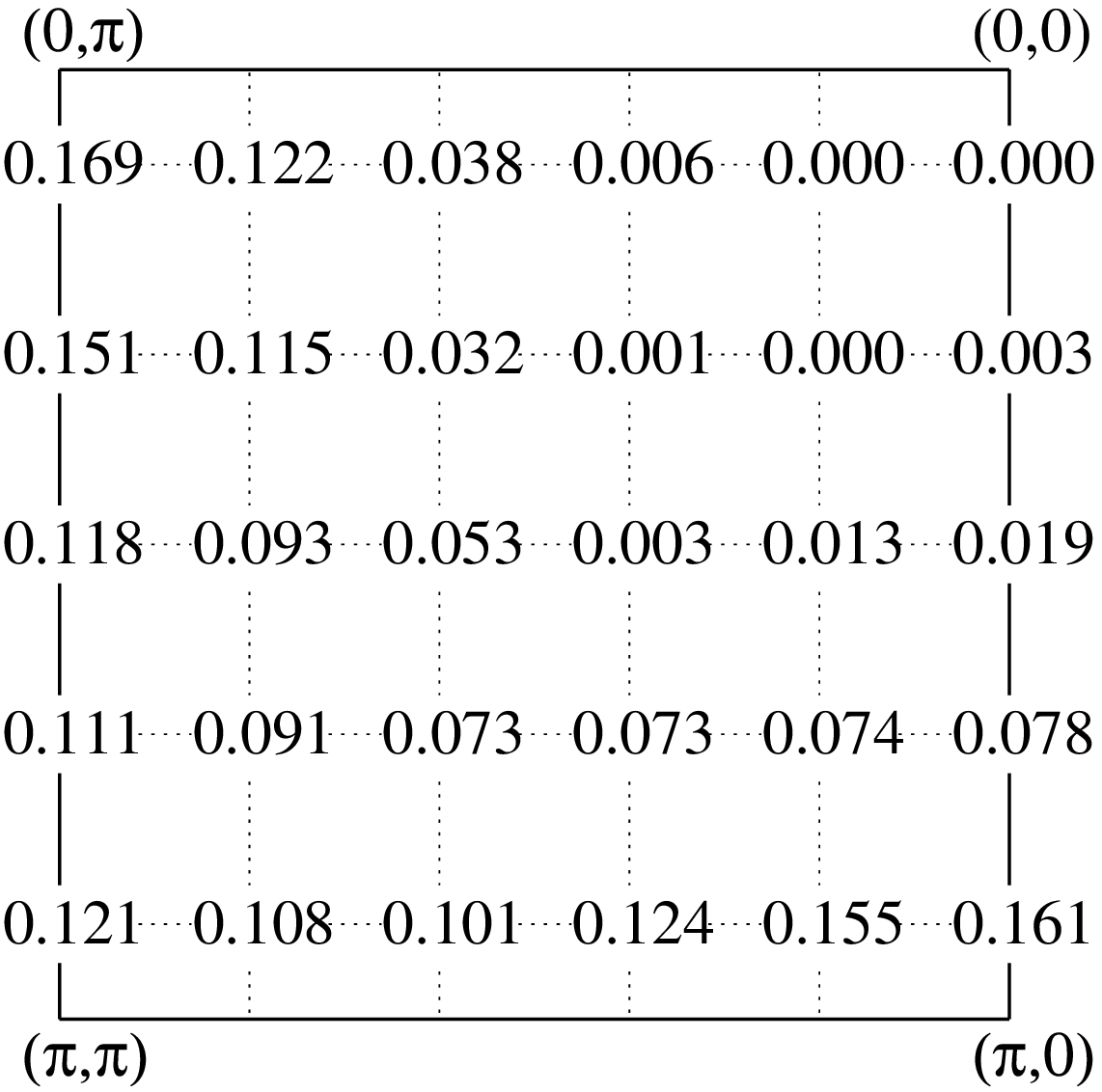}
\caption{ For one electron on 10 by 10 lattice, we plot $Z_{+,k}$ of
pBCSwf(left, $\chi=1$, $\Delta=0.55$) and pSDwf(right, $\chi=1$, $\Delta=0.55$,
$\frac{\tilde{\beta}_1}{\tilde{\beta}_0}=-0.5$,
$\frac{\tilde{\beta}_2}{\tilde{\beta}_0}=-0.3$,
$\frac{\tilde{\beta}_1}{\tilde{\beta}_0}=0.3$). By particle-hole symmetry, the
$Z_+$ of one electron pBCSwf is identical to the $Z_-$ of one hole pBCSwf
together with a $(\pi,\pi)$ momentum shift. pBCSwf has almost constant $Z_+$
along the direction from $(\pi,0)$ to $(0,\pi)$; while pSDwf has $Z_+$ suppressed
at nodal point and enhanced at anti-nodal point.  } \label{z-plot-ed}
\end{figure}

We did a variational search for the optimal variational parameters for pBCSwf
with longer range hoppings $\chi'$ and $\chi''$, and pSDwf with non-local
mixings.  In Table \ref{ed-energy} we compare the energy of pBCSwf and pSDwf
with 2 electron doped on 10 by 10 lattice. In Fig.\ref{z-plot-ed} we plot the
$Z_+$ map of pSDwf, one can see pSDwf has spectral weight of anti-dichotomy
shape.

\begin{table*} 
\begin{tabular}{|l|l|l|l|l|l|l|l|l|l|l|l|} \hline
wavefunction&$\frac{\Delta}{\chi}$&$\frac{\chi'}{\chi}$&$\frac{\chi''}{\chi}$&$\frac{\tilde{\beta}_1}{\tilde{\beta}_0}$&$\frac{\tilde{\beta}_2}{\tilde{\beta}_0}$&$\frac{\tilde{\beta}_3}{\tilde{\beta}_0}$&$\begin{array}{c}\mbox{total energy}\\
\mbox{per bond}\end{array}$&$\begin{array}{c}\la\vec S_i\cdot\vec S_{i+1}\ra\\\mbox{per bond}\end{array}$&$T_1$&$T_2$&$T_3$ \\ \hline
pBCSwf&0.55&0&0&0&0&0&-0.1884$\pm$0.0001&   -0.2977$\pm$0.0002 &
2.64$\pm$0.01 &   0.52$\pm$0.01  &  0.48$\pm$0.01\\ \hline
pBCSwf(optimal)&0.55&0.2&0.0&0&0&0&-0.1888$\pm$0.0001 &  -0.2964$\pm$0.0002 &
2.61$\pm$0.01   &0.70$\pm$0.02 &    0.20$\pm$0.02\\\hline
pSDwf(optimal)&0.55&0&0&-0.5 & -0.3 & 0.3 & -0.1910$\pm$0.0001 &
-0.2971$\pm$0.0002 &      2.57$\pm$0.01& 0.86$\pm$0.02&
-0.72$\pm$0.02\\\hline 
\end{tabular} \caption{ Two electrons on 10 by 10
lattice. $t=1$, $t'=-0.3$, $t''=0.2$ and $J=0.3$, and we mapped it into a
hole-doped model with $t=1$, $t'=0.3$, $t''=-0.2$ and $J=0.3$ with a
$(\pi,\pi)$ shift in momentum space. $T_1$, $T_2$ and $T_3$ stand for nearest neighbor hopping per hole $\frac{1}{N_h}\sum_{i,\delta=\pm\hat{x},\pm\hat{y}}\la c_i^{\dag}c_{i+\delta}\ra$, next nearest neighbor hopping per hole $\frac{1}{N_h}\sum_{i,\delta=\pm(\hat{x}+\hat{y}),\pm(\hat{x}-\hat{y})}\la c_i^{\dag}c_{i+\delta}\ra$ and third nearest neighbor hopping per hole $\frac{1}{N_h}\sum_{i,\delta=\pm2\hat{x},\pm2\hat{y}}\la c_i^{\dag}c_{i+\delta}\ra$ respectively. We compare pBCSwf of $d$-wave ansatz,
pBCSwf with longer range hoppings $\chi'$ and $\chi''$, and pSDwf with non-local
mixings. The best trial pSDwf has energy lowered by 1.2\% from the best trial
pBCSwf with longer range hoppings. And comparing the first line and the last line which have the same spin correlations, we find that the energy of a doped electron in pSDwf is $0.26t$ lower than that of an electron in pBCSwf.
} \label{ed-energy} 
\end{table*}

\begin{figure*}
\includegraphics[height=2.2in]
{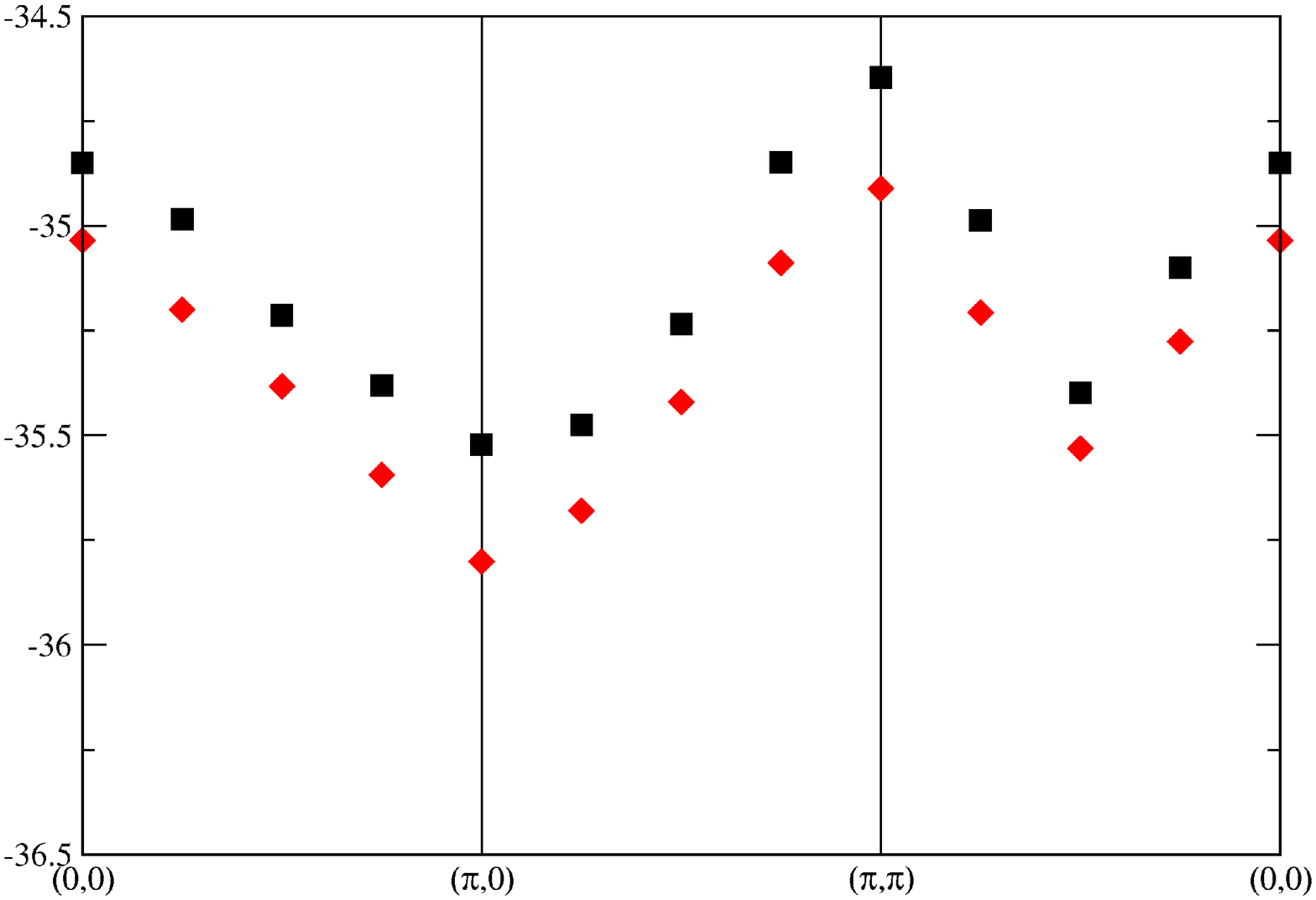}\;\;\;\;
\includegraphics[height=2.2in]
{excitation2-ed.eps}
\caption{
Quasi-particle spectrum for one electron on 10 by 10 lattice. $t=1$, $t'=-0.3$,
$t''=0.2$ and $J=0.3$ (one can map it into a hole-doped model with $t=1$,
$t'=0.3$, $t''=-0.2$ and $J=0.3$ with a $(\pi,\pi)$ shift in momentum space.).
The black square shows the spectrum of $d$-wave pBCSwf with $\chi=1$ and
$\Delta=0.55$, and the red diamond shows the spectrum of pSDwf with $\chi=1$,
$\Delta=0.55$, $\frac{\tilde{\beta}_1}{\tilde{\beta}_0}=-0.5$,
$\frac{\tilde{\beta}_2}{\tilde{\beta}_0}=-0.3$ and
$\frac{\tilde{\beta}_3}{\tilde{\beta}_0}=0.3$. One can see the first electron
doped to $(\pi,0)$.  The energy of a doped electron in pSDwf is $0.25t$ lower
than that of an electron in pBCSwf.
} 
\label{excitation-ed} 
\end{figure*}

In Fig.\ref{excitation-ed} we compare the energy dispersion of one-electron
quasi-particle excitations of pBCSwf and pSDwf. 
The energy of a doped electron in pSDwf is $0.25t$ lower
than that of an electron in pBCSwf.

\subsection{A prediction}

In hole-doped and electron-doped case, $t>0$ and $t'$ and $t''$ have opposite
signs, and as a result $Z_-$ develops strong $k$ dependence along diagonal
direction. What if $t'$ and $t''$ have the same sign? If both $t'>0$ and
$t''>0$, one expects that $\tilde{\beta}_2<0$ and $\tilde{\beta}_3<0$ to favor
longer range hoppings. But they drive the modulation of $Z_-$ in opposite ways.
As a result, one expects that for certain ratio of values of $t'>0$ and $t''>0$ of
order 1, their effects cancel and $Z_-$ remains constant along the diagonal
direction, but with an enhanced value of $Z_-$ than the case of pure $t$-$J$
model. Similarly for certain ratio of values of $t'<0$ and $t''<0$ of order 1,
$Z_-$ remains constant along the diagonal direction, but with a suppressed
value of $Z_-$ than the case of pure $t$-$J$ model. These predictions can be
checked by exact diagonalization.

\subsection{pBCSwf with longer range hoppings}\label{diff} 

One can view pSDwf as an improved pBCSwf. We choose the $d$-wave pairing
wavefunction $b(k)$ with only nearest hopping $\chi$ and pairing $\Delta$
parameters. Then $\beta_2$ and $\beta_3$ encode some second-neighbor and
third-neighbor correlations. The price to pay is to include more than one
Slater determinants in spin basis. One may naturally ask, suppose we insist
working on pBCSwf, if one puts in longer range hopping parameters like $\chi'$
and $\chi''$ in the pairing wavefunction $b(k)$, one also encodes some
second-neighbor and third-neighbor correlations, which may lower the
second-neighbor and third-neighbor hopping energies. But in this way one can
still work with a single Slater determinant.  If our pSDwf with no-local mixing
is physically similar to pBCSwf with longer range hoppings, why should one
bother to work with many Slater determinants? 

We want to emphasize that our pSDwf is physically different from pBCSwf even
after we include longer range hoppings $\chi'$ and $\chi''$.  We note that, in
the infinite-lattice limit with a few holes, the pBCSwf cannot have longer
range hoppings ({\it i.e.} $\chi'=\chi''=0$).  Otherwise we are considering
some other spin wavefunction instead of $d$-wave wavefunction, which will
increase the spin energy by a finite amount per site. Therefore $\chi'$ and
$\chi''$ have to vanish in low doping limit.  In contrast, for our pSDwf, the
the spin energy is not affected by finite $\beta_i$ in the zero doping limit.
Thus in the low doping limit, the spin energy is perturbed only slightly by a
finite $\beta_2$ and $\beta_3$. On the other hand a finite $\beta_2$ and
$\beta_3$ make the hopping energy much larger than that of pBCSwf.  So in the
infinite-lattice limit with a few holes, $\beta_i$ will be finite and the
energy of one hole will be lowered by a finite amount by turning on a finite
$\beta_i$.

Physically this means that $\beta_2$ and $\beta_3$ characterize the charge
correlations, while $\chi'$ and $\chi''$ characterize the spin correlations.
The above claim is supported by 2-hole system on larger lattice, i.e., by lower
the doping. In Table \ref{14by14} we list the energies of pBCSwf with longer
range hopping and pSDwf on 14 by 14 lattice. Comparing with Table \ref{thtb}
one can see the spin energy for pSDwf is lowered further than that for pBCSwf.

\begin{table*} 
\begin{tabular}{|l|l|l|l|l|l|l|l|l|l|l|l|} \hline
wavefunction&$\frac{\Delta}{\chi}$&$\frac{\chi'}{\chi}$&$\frac{\chi''}{\chi}$&$\frac{\tilde{\beta}_1}{\tilde{\beta}_0}$&$\frac{\tilde{\beta}_2}{\tilde{\beta}_0}$&$\frac{\tilde{\beta}_3}{\tilde{\beta}_0}$&$\begin{array}{c}\mbox{total energy}\\
\mbox{per bond}\end{array}$&$\begin{array}{c}\la\vec S_i\cdot\vec S_{i+1}\ra\\\mbox{per bond}\end{array}$&$T_1$&$T_2$&$T_3$ \\ \hline
pBCSwf&0.55&0&0&0&0&0&-0.1793$\pm$0.0001&   -0.3075$\pm$0.0002 &
2.65$\pm$0.02 &   0.43$\pm$0.02  &  0.65$\pm$0.02\\ \hline
pBCSwf&0.55&-0.4&0&0&0&0&-0.1796$\pm$0.0001 &  -0.3043$\pm$0.0002 &
2.66$\pm$0.02   &0.02$\pm$0.02 &    1.21$\pm$0.02\\\hline pSDwf&0.55&0&0&-0.3 &
0.3 & -0.1 & -0.1815$\pm$0.0001 &     -0.3058$\pm$0.0002 &      2.86$\pm$0.02&
-0.49$\pm$0.02&       0.87$\pm$0.02\\\hline 
\end{tabular} 
\caption{ Two holes
on 14 by 14 lattice. $t=1$, $t'=-0.3$, $t''=0.2$ and $J=0.3$. $T_1$, $T_2$ and $T_3$ stand for nearest neighbor hopping, next nearest neighbor hopping and third nearest neighbor hopping respectively. Although the
spin energy of pBCSwf with finite longer range hoppings $\chi'=-0.4$ is slightly
lower than that of pSDwf on 10 by 10 lattice, it is much higher on 14 by 14
lattice.  
} \label{14by14} 
\end{table*}

\begin{figure} 
\includegraphics[width=0.22\textwidth]{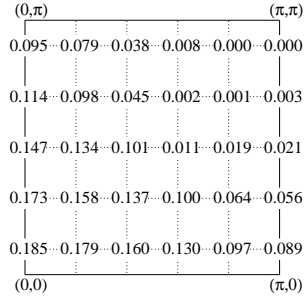}
\caption{For one hole on 10 by 10 lattice, we plot $Z_{-,k}$ of pBCSwf with
longer range hopping $\chi'=-0.4$ ($\chi=1$, $\Delta=0.55$).} \label{z-plot-chip}
\end{figure}

Another way to see that these two wavefunctions are different is by calculating
$Z_-$. Numerical results show that pSDwf has dichotomy whereas pBCSwf does not.
Actually on the mean-field level, a negative $\chi'$ and/or positive $\chi''$
even make the $Z_-$ larger on the anti-nodal point than on the nodal point.
After projection, we observe that $Z_-$ still remains almost constant along the
diagonal direction for pBCSwf with longer range hoppings. In Fig.\ref{z-plot-chip}
we plot the $Z_-$ map of pBCSwf with longer range hopping $\chi'=-0.4$.

\section{Conclusion}

In this paper we studied a new type of variational wavefunction, pSDwf. It can
be viewed as an improved pBCSwf, and the improvement is that pSDwf correctly
characterizes the charge dynamics and the correlation between the doped
holes/electrons and the nearby spins. This physics was missed by the previous
pBCSwf.  As a result, pSDwf correctly reproduces the dichotomy of hole-doped
and electron-doped Mott insulator. 

In pSDwf, we introduced two types of fermions, spinon $f$ and dopon $d$.
Spinons $f$ carry spin but no charge. They form a $d$-wave paired state that
describes the spin liquid background. Dopons $d$ carry both spin and charge and
correspond to a bare doped hole.  The mixing between spinons and dopons
described by $\beta_0$, $\beta_1$, $\beta_2$ and $\beta_3$ leads to a $d$-wave
superconducting state.  The charge dynamics (such as electron spectral
function) is determined by those mixings.  $\beta_0$ is the on-site mixing (or
local mixing), and $\beta_1$, $\beta_2$ and $\beta_3$ are non-local mixings
corresponding to mixing with first, second and third neighbors respectively. If
pSDwf has only local mixing, it is identical to pBCSwf. With non-local mixings,
pSDwf corresponds to pBCSwf with hopping terms acting on it.  Therefore the
wavefunction develops finite non-local mixings to lower the hopping energies.
In particular, for the hole-doped case, to lower $t'$ and $t''$ energies,
the mixing is described by $\beta_2>0$ and $\beta_3<0$.

The pSDwf can also be obtained by projecting the spinon-dopon mean-field
wavefunction into the physical subspace.  Therefore, one expects that some
properties of pSDwf can be understood from the mean-field theory.  In the
mean-field theory, it is clear that the modulation of $Z_{-,k}$ in $k$ space is
controlled by the non-local mixings.  Our numerical calculation of $Z_{-,k}$
shows that the above mean-field result is valid even for the projected wave
function.  We find that $\beta_2>0$ and $\beta_3<0$ give exactly the dichotomy
of $Z_{-,k}$ observed in the hole doped samples.  Because $\beta_2>0$ and
$\beta_3<0$ are driven by $t'$ and $t''$, the dichotomy is also driven by $t'$
and $t''$. Thus to lower the hopping energy, the spectral weight is suppressed
in some region in $k$-space. This result conflicts a naive guess: to lower the
hopping energy, the excitation should be more quasi-particle like.  We also
predict that the dichotomy will go away if $t'$ and $t''$ have the same sign
and similar magnitude. In summary, we found a mean-field theory and the
associated trial wavefunction capturing the dichotomy physics.

Traditionally, in projected wavefunction variational approach, for example
pBCSwf, people use wavefunctions which in real space correspond to a single
Slater determinant. The reason to do so is simply to make the computation
easier. Our study shows what kinds of important physics that may be missed by
doing so. In real space, the pSDwf is sum of $(1+n_{shift})^{n_{hole}}$ number
of Slater determinants, because each hole can either do not hop, or hop into
one of $n_{shift}$ sites. So our calculation is limited to few-hole cases.
However, the idea of introducing many Slater determinant is quite general. For
example, one can study another improved pBCSwf, which allows each hole to hop
once but forbids two holes hopping together, therefore the number of Slater
determinant is $(1+n_{shift})n_{hole}$ and many-hole cases are computationally
achievable. This new improved pBCSwf is the first order approximation of pSDwf
and remains to be studied. For a long time there is a puzzle that doped
Mott-insulator (\ie the spin disordered metallic state) seems to be
energetically favorable only at high doping $x>0.3$. For $x<0.3$ the doped spin
density wave state have a lower energy. 
Our pSDwf may push this limit down to low doping which agrees with experiments
better.  This is because that including many Slater determinants can lower the
energy per hole by a significant amount (about $0.4t$).

As we have stressed, pSDwf provides a better description of spin-charge
correlation, or more precisely, the spin configuration near a doped hole.  This
allows us to reproduce the dichotomy in quasiparticle spectral weights observed
in experiments.  The next question is whether the better understanding of the
spin-charge correlation can lead to new experimental predictions.  In the
following, we will describe one such prediction in quasi-particle current
distribution.

We know that a finite supercurrent $\v J_s$ shifts the superconducting
quasiparticle dispersion $E_{\v k}$. To the linear order in $\v J_s$, we have
\begin{equation*}
 E_{\v k}(\v A)=E_{\v k}(0)+c^{-1}\v j_{\v k}\cdot \v A ,
\end{equation*}
where $c$ is the speed of light and we have introduced the vector potential $\v
A$ to represent the supercurrent: $\v J_s= \frac{n_s e^2}{m c}\v A$.  $\v j_{\v
k}$ is a very important function that characterizes how excited quasiparticles
affect superfluid density $\rho_S$.  We call $\v j_{\v k}$ quasiparticle
current.  According to the BCS theory 
\begin{equation}
\label{jqBCS}
\v j_{\v k}=e\frac{\prt \epsilon_{\v k}}{\prt \v k}=e\v v_\text{normal},
\end{equation}
where $\epsilon_{\v k}$ is the normal state dispersion 
which is roughly given by $\epsilon_{\v k}=-2t[\cos(k_x)+\cos(k_y)]$.

The previous study\cite{8880404} of quasi-particle current for pBCSwf shows
that the quasi-particle current is roughly given by the BCS result
(\ref{jqBCS}) scaled down by a factor $\alpha$.  Such a quasi-particle current
has a smooth distribution in $k$-space.  Here we would like to stress that
since the charge dynamics is not capture well by the pBCSwf, the above result
from pBCSwf may not be reliable.  We expect that the quasi-particle current
of pSDwf should has a strong $k$-dependence, \ie a large quasi-particle current
near the nodal point where $Z_k$ is large and small quasi-particle current near
the anti-nodal point where $Z_k$ is small.  Such a  quasi-particle current
distribution may explain the temperature dependence of superfluid
density \cite{liang:117001}.

Indeed, the mean-field spinon-dopon approach does give rise to a very different
quasi-particle current distribution which roughly follows $Z_k$.  For more
detailed study in this direction and possible experimental tests, see Ref.
\cite{TWA}.

We would like to thank Cody Nave, Sung-Sik Lee, P. A. Lee for helpful
discussions.  This work was supported by NSF grant No.  DMR-0433632

\appendix 

\section{A simple algorithm to do local projection}\label{LP} 

Suppose
the wavefunction before projection is the ground state of some fermionic
quadratic Hamiltonian. One can always diagonalize the Hamiltonian so that all
two-point correlation functions of fermion operators can be calculated exactly.
For our SDwf, that means quantities like $\la f_i^{\dag} f_j\ra$, $\la
d_i^{\dag} f_j\ra$,$\la d_i^{\dag} d_j\ra$... can be calculated. 

Projection is supposed to remove the unphysical states. For a site $i$, the
following operator removes the unphysical states.  
\begin{align}
P_i=&n_{f,i}(n_{f,i}-2)(\frac{1}{2}n_{d,i}^2-\frac{1}{2}n_{d,i}+1)\notag\\
&\cdot(1-\frac{1}{2}n_{d,i}(\vec
S_{f,i}+\vec S_{d,i})^2).
\end{align} 
It obviously ensures that
$n_{f,i}=1$, $n_{d,i}\neq 2$ and $f$ and $d$ fermions form local singlet. To
calculate energy, we do local projection on the relevant sites. For example, to
calculate the $J$ term energy, one actually calculates 
\begin{align} \langle
\vec S_i \cdot \vec S_j\rangle_{prj}=\frac{\langle P_iP_j (\vec
S_{f,i}+\vec S_{d,i}) \cdot (\vec
S_{f,j}+\vec S_{d,j})
P_iP_j\rangle}{\langle P_i P_j \rangle}.
\end{align} 
The denominator accounts for
the wavefunction normalization due to projection.  One can write operators $P_i
\vec S_i \cdot \vec S_j P_j$ and $P_i \vec S_i \cdot \vec S_j P_j$ in terms of
fermion operators. By Wick's theorem, the expectation values of these operators
reduce to a sum of products of fermion two-point correlation functions, which
are known. Similarly for $t$ term energy, one calculates for example,
\begin{align} \la
c_{i\uparrow}^{\dag}c_{j\uparrow}\ra_{prj}=\frac{\langle P_iP_j f_{i\uparrow}^{\dag}  h_{j}^{\dag} h_{i}f_{j\uparrow} P_iP_j\rangle}{\langle
P_i P_j \rangle},
\end{align}
where $h_{i}^{\dag}=\frac{1}{\sqrt{2}}(f_{i\uparrow}^{\dag}d_{i\downarrow}^{\dag}-f_{i\downarrow}^{\dag}d_{i\uparrow}^{\dag})$ is the operator that creates a hole at site $i$.

One may ask whether we can do local projections on more and more sites, then
the result will be closer and closer to the one of full projection.
Unfortunately this cannot be done, because the number of terms in the summation
when we expand $P_{i1}P_{i2}\ldots P_{in}$ increases exponentially fast as we
increase $n$. Therefore we are limited to few sites. The above method can only
be viewed as some renormalized mean-field approach.

\bibliographystyle{apsrev}
\bibliography{/home/ranying/downloads/reference/simplifiedying,/home/ranying/downloads/reference/wen}

\end{document}